\documentclass[letterpaper,twocolumn,10pt, hyphens,obeyspaces,spaces]{article}

\RequirePackage[2020-02-02]{latexrelease}

\usepackage{usenix-2020-09}

\usepackage{titlesec}
\titlespacing{\section}{0pt}{6pt plus 2pt minus 2pt}{4pt plus 2pt minus 2pt}
\titlespacing{\subsection}{0pt}{6pt plus 2pt minus 2pt}{4pt plus 2pt minus 2pt}
\titlespacing{\subsubsection}{0pt}{3pt plus 2pt minus 2pt}{2pt plus 1pt minus 1pt}
\titlespacing{\paragraph}{0pt}{\parskip}{-\parskip}

\PassOptionsToPackage{usenames,dvipsnames}{xcolor}
\usepackage{url}
\usepackage[utf8]{inputenc}
\usepackage{array, multirow}
\newcolumntype{P}[1]{>{\centering\arraybackslash}p{#1}}
\usepackage{tikz}
\usepackage{makecell}
\usepackage{subcaption}
\usepackage{flushend}
\usepackage{fancyhdr}
\usepackage{algorithm2e}
\usepackage{listings}
\usepackage{cprotect}
\usepackage{environ}
\usepackage{authblk}
\usepackage{amsmath}
\usepackage{mathtools}
\usepackage{svg}
\usepackage{xcolor}
\usepackage{graphicx}
\usepackage{tikz}
\usetikzlibrary{trees}                          % For building tree like structures
\usetikzlibrary{arrows}

\newcommand{\updated}[1]{\color{black} #1}

\usepackage{graphicx}
\usepackage{caption}

\renewcommand{\arraystretch}{1.25}
\usepackage{pifont}
\newcommand{\cmark}{\ding{51}} % tick  aka yes
\newcommand{\xmark}{\ding{55}} % cross aka no

\newcommand{\ignore}[1]{}
\newcolumntype{P}[1]{>{\centering\arraybackslash}p{#1}}
\usepackage{tikz}
\usepackage{amsmath}
\usepackage{amssymb}
\newcommand{\stt}[1]{{\small\path{#1}}}

\newcolumntype{H}{>{\setbox0=\hbox\bgroup}c<{\egroup}@{\hspace*{-\tabcolsep}}}

\begin{document}
\pagenumbering{gobble} 

\title{Keep Your Friends Close, but Your Routeservers Closer:\\Insights into RPKI Validation in the Internet}

\author[*$\S$]{Tomas Hlavacek}
\author[*$\S\dag$]{Haya Shulman}
\author[*$\S\dag$]{Niklas Vogel}
\author[*$\S\ddag$]{Michael Waidner}
\affil[$\S$]{National Research Center for Applied Cybersecurity ATHENE}
\affil[*]{Fraunhofer Institute for Secure Information Technology SIT}
\affil[$\ddag$]{Technische Universität Darmstadt}
\affil[$\dag$]{Goethe-Universität Frankfurt}

\maketitle

\begin{abstract}
IP prefix hijacks allow adversaries to redirect and intercept traffic, posing a threat to the stability and security of the Internet. To prevent prefix hijacks, networks should deploy RPKI and filter bogus BGP announcements with invalid routes.

In this work we evaluate the impact of RPKI deployments on the security and resilience of the Internet. We aim to understand which networks filter invalid routes and how effective that filtering is in blocking prefix hijacks. We extend previous data acquisition and analysis methodologies to obtain more accurate identification of networks that filter invalid routes with RPKI. We find that more than 27\% of networks enforce RPKI filtering and show for the first time that deployments follow the business incentives of inter-domain routing: providers have an increased motivation to filter in order to avoid losing customers' traffic.

Analyzing the effectiveness of RPKI, we find that the current trend to deploy RPKI on routeservers of Internet Exchange Points (IXPs) only provides a localized protection against hijacks but has negligible impact on preventing their spread globally. In contrast, we show that RPKI filtering in Tier-1 providers greatly benefits the security of the Internet as it limits the spread of hijacks to a localized scope. Based on our observations, we provide recommendations on the future roadmap of RPKI deployment.

We make our datasets\footnote{\url{https://sit4.me/rpki}} available for public use.
\end{abstract}

\section{Introduction}

{\bf BGP prefix hijacks.} The Internet consists of Autonomous Systems (ASes) connected with the Border Gateway Protocol (BGP) [RFC1105]. Routers in each AS send BGP announcements to notify other networks how to reach IP addresses within prefixes that they own. BGP announcements are not authenticated, hence border routers can issue announcements claiming to originate {\em any} Internet prefix. Such bogus announcements can be a result of benign misconfigurations or malicious attacks \cite{china:telecom,fb:out,turkey:hijack}. ASes accepting bogus announcements send the traffic via invalid paths to the hijacker instead of the legitimate destination \cite{bellovin1989security}. BGP prefix hijacks allow adversaries to intercept, manipulate, and blackhole communication \cite{ballani2007study,vervier2015mind}. 

{\bf Filtering invalid routes with RPKI.} To prevent prefix hijacks, the IETF standardized the Resource Public Key Infrastructure (RPKI) [RFC6480]. The RPKI authenticates ownership over prefixes by binding prefixes to AS numbers (ASNs) and to public keys, creating Route Origin Authorizations (ROAs). The ROAs are stored in RPKI publication points. To filter bogus announcements, ASes should enforce Route Origin Validation (ROV): use relying party validators to periodically fetch and validate ROAs, and use these validation results in border routers to make routing decisions in BGP. 
Although ROV is critical for preventing hijacks, the deployment of ROV has only seen a moderate pace after its introduction in 2013. In recent years, the deployment took off with the adoption of ROV by Internet Exchange Points (IXPs) and large providers. {\em Our goal is to understand how ROV in different network types affects propagation of invalid paths and how effective ROV deployments are in blocking hijacks.}

{\bf Measurements of ROV.} Due to its significant role to Internet security, understanding the fraction of ASes that enforce ROV poses an important research question. In this work we measure ROV via a combination of control and data-plane measurements using RIPE Atlas\footnote{\url{atlas.ripe.net}}, similarly to \cite{hlavacek2018practical,rodday2021revisiting}.
 
We create invalid ROAs that conflict with the BGP announcements for our prefixes and hence appear like prefix hijacks.
To identify ASes that change their routing to our prefixes, we inspect control-plane paths in the global BGP routing table and measure the data-plane routes that the traffic takes to our prefixes. ASes whose routing to our invalid prefixes is not affected do not enforce ROV. 
In contrast to other approaches for measuring ROV, which we discuss in Related Work, Section \ref{sc:works}, this approach provides the best coverage of the Internet, is scalable, and does not require volunteers. We also improve the data analysis and acquisition of previous work to eliminate random routing events, therefore reducing the high fraction of false negatives/positives in previous research \cite{hlavacek2018practical,rodday2021revisiting}.

{\bf Business incentives for ROV enforcement.} In addition to the improved methodology, we also characterize the ASes that enforce ROV in our measurements according to their type and size. Through our analysis we find a correlation between the business model of ASes and ROV enforcement, and show that this correlation is aligned with the business incentives of BGP: 

{\em Large ASes, Internet Service Providers (ISPs) and IXPs have increased motivation to enforce ROV, since they get paid for providing connectivity services. Consequently, when a prefix is hijacked, they lose traffic and corresponding payment. Prefix hijacks affect their business model}. In contrast, stub-ASes do not provide upstream connectivity to other networks, hence do not have a business incentive to enforce ROV themselves. 

% hijack vs propagation
{\bf Effectiveness in blocking invalid paths.} Our goal is not to merely understand if hijacks are possible, e.g. like \cite{gilad2017we}, but
to gain insights into how far the invalid routes can reach, the scope of the affected networks, the impact of ROV on reachability of ASes, which parts of the Internet are not protected, and which networks play a central role in providing global protection against hijacks. We evaluate the effectiveness of current ROV deployments through analysis of the propagation of invalid routes across different network types. Although there are suggestions that ROV at routeservers of IXPs provides an effective defence against prefix hijacks \cite{reuter2018towards,rodday2021revisiting}, we show for the first time that IXPs do not block global propagation of invalid routes. Routeservers at IXPs perform control-plane functions interconnecting border routers of customer ASes, to manage peerings and to guarantee protection to the customers against BGP hijacks by dropping invalid routes via ROV. Outsourcing the management of peerings and blocking of hijacks with ROV to the IXP made the routeservers extremely popular. We show that IXPs cannot prevent leakage of invalid paths globally because they do not have control over the traffic routed through their IP space over direct sessions. In fact, we find that the average direct peering sessions in the top five IXPs propagate 3.4x more paths than sessions over a routeserver, inevitably leaking invalid routes. In contrast, we find that ROV enforcement in Tier-1 providers is most effective in blocking global propagation of invalid routes.

{\bf Research questions.}  In our research we aim to answer the following questions. 

$\bullet$ The enforcement of ROV is changing at a rapid pace. What is the current ROV deployment rate in the Internet?

$\bullet$ What limitations do existing methodologies for measuring ROV have, what measurement bias do they introduce, and how can they be improved? 

$\bullet$ What are the differences between control and data plane methodologies, what is the overlap, and what are the factors that cause the differences?

$\bullet$ Are there differences in ROV enforcement between different networks and geo-locations?

$\bullet$ In which networks is ROV enforcement most effective for blocking hijacks?

{\bf Ethical considerations.} In order to identify ASes that enforce ROV, we carry out active BGP prefix hijacks in the global Internet and measure which ASes accept the routes in our hijacking announcements. Our experiments are ethical; we hijack {\em only} the prefixes that we own. Our experiments do not introduce additional load on other networks.

{\bf Contributions.} Conceptually, our research shows that IXPs play a much smaller role in blocking invalid routes than {\updated indicated by previous research \cite{rodday2021revisiting}, which concluded that most ROV enforcement is performed in the IXPs}. 
In contrast, our analysis demonstrates that ROV in Tier-1 providers significantly reduces the global propagation of invalid routes limiting the spread to a localized scope. We find that current ROV deployments do not provide sufficient protection against prefix hijacks and are not resilient to attacks and failures. Our technical contributions are:

$\bullet$ {\em Improved ROV measurements:} We improve the data acquisition and extraction processes used in previous ROV measurement studies \cite{hlavacek2018practical,rodday2021revisiting}. %Our data acquisition is more scalable, provides better coverage and eliminates random routing events. 
For data analysis, we introduce an AS classification scheme and divergence points into our methodology; both significantly reduce false positives and negatives inherent in previous measurements \cite{hlavacek2018practical,rodday2021revisiting}. We provide our dataset and instructions for reproducing our study at \url{https://sit4.me/rpki}.

$\bullet$ {\em Invalid paths over IXPs:} We performed the first study of the effectiveness of routeserver-ROV in blocking invalid paths. Our measurements covered 159 IXPs, including the largest European IXPs, and found route leaks over them. 

$\bullet$ {\em Propagation of invalid routes:} We develop the first graph-based analysis of ROV effectiveness on limiting the propagation of invalid paths. Using our analysis, we evaluate the outreach of invalid paths on the Internet and identify networks whose ROV filtering provides effective global protection.  

{\bf Organization.} We review RPKI in Section \ref{sc:overview} and compare our research to Related Work in Section \ref{sc:works}. We introduce our ROV-measurement methodology in Section \ref{sc:method}. The setup and execution of the experiments are explained in Section \ref{sc:measurements}, and the results of ROV measurement are presented in Section \ref{sc:rov}. We quantify the invalid paths that traverse the IXPs in Section \ref{sc:routeservers}. In Section \ref{sc:invalid:paths}, we analyze the effectiveness of ROV filtering on blocking the propagation of invalid routes in the Internet. We conclude our research in Section \ref{sc:conclusions}.

\section{Overview of RPKI}\label{sc:overview}

RPKI provides authenticated prefix ownership information, which routers can use for making routing decisions.

{\bf RPKI objects.} To authorize their network resources, ASes can create resource certificates that bind their resources to a public key contained inside a Route Origin Authorization. The ROA is signed with the certificate of a Certificate Authority (CA). RPKI objects are published in RPKI repositories hosted on publication points. The publication points are operated either in a hosted mode by one of the Regional Internet Registries (RIRs) or in a delegated mode by a Local Internet Registry (LIR). An RPKI repository keeps a finite set of signed ROAs and additionally contains signed certificates (which point to children publication points), certificate revocation lists (CRLs), and manifests.

{\bf Traversal of RPKI repositories.} The validation of RPKI objects is performed with a relying party software, which contains hardcoded Trust Anchor Locators (TALs) to the root certificates of the RIRs. Each of the five RIRs operates its own RPKI trust anchor certificate and repository. During the validation, a relying party contacts every root repository known to it and downloads RPKI objects from every publication point it finds. The RPKI objects are fetched from RPKI repositories over RRDP or rsync protocols. After downloading the objects, a relying party performs cryptographic validation, which produces a list of tuples (AS, ROA prefix, prefix length) called Validated ROA Payloads (VRPs). The VRPs are stored in a local cache. 

{\bf Route Origin Validation.} The BGP border routers of an AS retrieve the VRPs from their relying party's cache over the `RPKI to Router Protocol' (RTR) [RFC8210]. The routers use the VRPs to validate incoming BGP announcements with Route Origin Validation. A router checks if the IP prefix block in the BGP announcement and the VRP IP prefix block are identical for the length specified by the VRP IP prefix length [RFC6811]. If the IP prefix in the announcement is covered by any VRP entry, the router checks if the BGP origin AS in the announcement matches the VRP AS for that prefix. Matching values result in the conclusion that the announcement is valid. In contrast, if any VRP covers the prefix in the BGP announcement, but the entry does not match the origin AS, then the announcement is invalid. The validation status is considered unknown if the BGP announcement is not covered by any VRP entry. 
\section{Related Work}\label{sc:works}
\begin{table*}[t!]
\renewcommand{\arraystretch}{0.6}
    \centering
    \footnotesize
\begin{tabular}{l|c|c|c|c|c|c|c|c}
\textbf{Research} & \textbf{Year} & \textbf{Control-plane} & \textbf{Data-plane} & \textbf{Methodology with} & \textbf{Clients} & \textbf{>1 Target} & \textbf{Rate of} & \textbf{\#ASes}\\
 &  & \textbf{} & \textbf{} & \textbf{Divergence} & & \textbf{AS} & \textbf{ROV-ASes} \\\hline \hline
This work & 2022 & \cmark  & Traceroute/Atlas & \cmark & \xmark & \cmark & 27\% & 2.4K\\ \hline
Cloudflare \cite{cloudflare} & 2022 & \xmark & HTTP & \xmark & Volunteers & \xmark & 30\% & 380 \\ \hline
Rodday et al. \cite{rodday2021revisiting} & 2021 & \cmark  & Traceroute/Atlas & \xmark & \xmark & \xmark & 0.6\% & 3.6K\\ \hline 
APNIC \cite{apnic} & 2021 & \cmark & HTTP & \xmark & Ad-network & \xmark & 25\% & 25K \\ \hline
Testart et al. \cite{testart2020filter} & 2020 & RouteViews & \xmark & \xmark & \xmark & \xmark & 11\% & 21 \\ \hline
Hlavacek et al. \cite{hlavacek2018practical} & 2018 & \cmark & Traceroute/Atlas & \xmark & \xmark & \cmark & 0.5\% & 296\\ \hline
Gilad et al. \cite{gilad2017we} & 2017 & \cmark & \xmark & \xmark & \xmark & \xmark & 3\% & 100 \\ %\hline
\end{tabular}
\vspace{-7pt}
\caption{Measurements of ROV: characteristics of our and previous work.}
\vspace{-10pt}
\label{tab:comparison}
\end{table*}
Practical impact of BGP prefix hijacks has been extensively explored \cite{DBLP:conf/uss/Birge-LeeSERM18,DBLP:journals/corr/abs-2004-09063} and real-world hijack incidents \cite{china:telecom,mitm:threat,turkey:hijack,indosat:hijack} confirmed the projected assessment of the research. The awareness to prefix hijacks creates a strong motivation to understand the deployment of RPKI and to obtain insights into the effectiveness of ROV. Previous measurements studied related aspects, such as the prevalence of invalid ROA objects caused by benign misconfigurations \cite{chung2019rpki,DBLP:conf/esorics/HlavacekSW22}, the impact of the Domain Name System on the resilience of RPKI \cite{DBLP:conf/ccs/HlavacekJMSW22} or downgrade attacks against RPKI \cite{usenix-stalloris-21,DBLP:conf/ccs/MirditaSW22}. In this work we explore the effectiveness of ROV. We next put our research in the context of related work on measurements of ROV.

{\bf Effectiveness of ROV.} Previous work provided a theoretical upper bound on the feasibility of hijacks \cite{gilad2017we,hlavacek2020disco}. 
This was done by simulating success of any Internet AS to hijack any prefix assuming a varying fraction of ROV-enforcing ASes. Such simulations do not consider the data-plane paths that actual traffic takes and do not use the real ASes that de facto enforce ROV, but just assume a fraction of ROV enforcement. Therefore the theoretical bound does not reflect a realistic attack surface.
In addition to not reflecting practical factors relevant to success of hijacks, the simulations do not answer questions related to the effectiveness of ROV in blocking propagation of hijacks and to the affected networks. For instance, not all hijacks have equal impact and hijacking a Tier-1 provider also redirects the traffic of all its customers. Our goal is not only to understand if hijacks are feasible, but also to infer which and how many networks are affected by the hijacks and by the ROV filtering. We do this by analyzing the date-plane paths that traffic takes, and the impact of ROV on the Internet graph of networks. We use the observations from our analysis to derive future directions that deployment of ROV should take to reach optimal protection of the Internet.

{\bf Approaches for measuring ROV.} 
The first global measurement of ROV enforcement was carried out in 2017 \cite{gilad2017we} (listed in Table \ref{tab:comparison}). The study monitored the propagation of invalid BGP announcements in public BGP collectors and found 100 ROV-enforcing ASes. In their experiment, Gilad et al. \cite{gilad2017we} passively monitored ASes that originated valid and invalid BGP announcements, and then collected ASes that were on the paths towards the valid prefix, but not on the paths towards the invalid prefix. Those ASes were classified as ROV-enforcing. However, the measurements had high false positives and false negatives rates since they used invalid BGP announcements of other ASes, which they did not control. This also limited the coverage of the experiment. The methodology of \cite{gilad2017we} was improved with a controlled experiment in the control plane by \cite{hlavacek2018practical}, which monitored propagation of invalid announcements in public collectors and used active probes over RIPE Atlas. The study of \cite{hlavacek2018practical} found 296 ROV-enforcing ASes. A subsequent study in 2020 \cite{testart2020filter} passively analyzed the historical data from RouteViews\footnote{\url{http://www.routeviews.org}} to identify changes in routing behavior, finding 21 ROV-enforcing ASes. Since these measurements were performed using a limited number of collectors (less than 0.01\%) the results were not representative of the entire Internet. Increasing the coverage is imperative for collecting representative data. In 2021 \cite{rodday2021revisiting} did an ROV study with a methodology of \cite{hlavacek2018practical} using 5537 probes in 3694 origin ASes. 

In our work, we combine control and data-plane measurements similarly to \cite{hlavacek2018practical,rodday2021revisiting}. In contrast to \cite{hlavacek2018practical,rodday2021revisiting}, which used an invalid ROA conflicting with a BGP announcement to infer ROV enforcement, we alternate between valid and invalid BGP announcements, which has faster convergence time than changes in ROAs. Alternating between valid and invalid events using two prefixes during data acquisition further allows us to eliminate random routing events, such as events in which an AS uses traffic engineering that complies with the ROA validity, which may be misinterpreted as ROV. This alternation reduces false positives. In addition, the previous method does not scale since it adds a large number of false negatives that lack sufficient evidence for ROV enforcement. We explain the issues with false-positives and false-negatives in previous work when we derive our methodology from previous approaches in Section \ref{sc:method}. In our ROV measurements, we greatly reduce the number of false-negatives and thus provide a more realistic view of real-world ROV enforcement. During data analysis, we apply a path-aware methodology that uses a new metric, divergence points, to reduce false positives. Further, we introduce an AS classification scheme to differentiate ASes that actively enforce ROV from ASes with only passive protection in an upstream ROV. Our approach shows a much higher rate of ROV deployment in the Internet than previously found in \cite{hlavacek2018practical,rodday2021revisiting}, including extensive evidence for enforcement in 9 of the 15 Tier-1 providers. We show that the higher rates of ROV enforcement are related to the improvements in our methodology and the continuous increase of ROV enforcement rate over time.

 In 2023, an online service called RoVista\footnote{\url{https://rovista.netsecurelab.org/}} was set up for reporting ROV enforcement. Similarly to \cite{gilad2017we}, RoVista uses an uncontrolled control-plane experiment, utilizing ASes with invalid BGP announcements and probing the reachability to the invalid prefixes from other systems in the Internet. Since they use invalid routes that happen to be announced by other ASes, their coverage is limited; currently, as they point out, only 1\% of prefixes are RPKI invalid. Additionally, they have the same downsides as uncontrolled experiments like \cite{gilad2017we}, which include a high rate of false positives. For instance, ASes might appear to behave like ROV filtering networks because of other (non-ROV) mechanisms. A more significant issue with RoVista is the usage of IPID side channel to identify ASes that follow invalid routes. There are three problems with IPID side channels. First, globally incrementing IPID has been gradually phased out in operating systems. Previous work \cite{pearce2017augur,dai2021smap,DBLP:conf/dsn/ShulmanZ21} showed that very few hosts ($\sim$16\%) had globally incrementing IPID counters and that some hosts with globally incrementing counters set their values to 0 when packets are too small to be fragmented. RoVista additionally requires that multiple ASes have at least ten hosts with globally incrementing IPID. The authors do not provide the methodology and measurement details on how many ASes have at least ten hosts with globally incrementing IPID counters. Since previous work showed that the number of hosts with global incrementing IPID is small and is further shrinking, the approach has limited scalability. 

 Second, measurements of IPID incur a lot of noise due to communication from other hosts, failures, and traffic fluctuations. Simple applications that use IPID, such as indirect measurements of idle port scans with NMap, exhibit high failure rates in dynamic Internet environments with over 10\% failures in scans of even completely idle hosts. These measurements were slightly improved with anti-noise techniques used by \cite{zhang2018onis}. Using the IPID side channel to measure ROV enforcement appears to be much more challenging than just checking if a port is open. In particular, there is a large time interval between the probing of the IPID value and the time that the BGP announcements converge and the routes are updated. The lack of visibility into the exact time when the route change is accepted makes it impossible to approximate the probe time of the IPID value. This is expected to introduce an immense amount of noise into the measurements, producing many false negatives and positives, making it impossible to derive a conclusion on ROV enforcement. The authors do not explain how they deal with that noise. Finally, the traffic volume to the ASes that announce the invalid prefixes would become prohibitive in the presented methodology as all the tested networks are required to send traffic to these ASes from ten of their hosts, resulting in regular traffic from 280.000 hosts. In total, they send traffic from these hosts to 47 ASes that announce invalid prefixes. The lack of methodology details of the measurements done by RoVista makes it impossible to compare our study to theirs and to understand the correctness or effectiveness of their approach. 

A completely different approach was taken by the Cloudflare project \footnote{https://isbgpsafeyet.com/}, which is a community-driven effort to summarize ROV implementation of large providers. The project provides a webpage to test ROV enforcement of providers by probing the reaction of the client to a valid and invalid announcement. If a client can reach the valid announcement and not the invalid one, they conclude that the provider of the client enforces ROV. The user can then contribute to the project over a GitHub page and update the status of its provider in the dataset. Our measurements show that while this approach may be sensible for a rough overview of large ROV-enforcing networks, it does not scale to an accurate representation of ROV enforcement in the Internet. 
In contrast to previously described approaches, Cloudflare requires coordination and support of volunteers to measure ROV enforcement in the networks of the users, which proves problematic.
First, the dataset is small, with only 380 ASes. Second, smaller ASes with fewer clients are less likely to have a user that contributes, and thus the dataset is biased towards the largest providers. 
We also find false positives in the results. Since the measurement methodology relies on two announcements picked up by a single vantage point, it leads to errors in cases where the ROV in an upstream provider filters the invalid announcement. The provider of the user is mistakenly classified as ROV-enforcing. 
This approach thus not only limits the applicability of ROV measurements in the global Internet but also introduces a bias to the results. We evaluated the Cloudflare dataset and identified differences and errors in the classification, which we discuss in Section \ref{subsec:vali}.

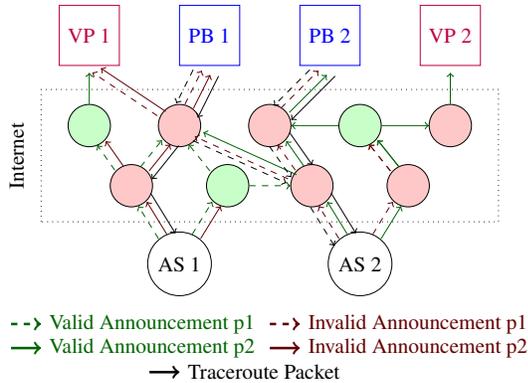
\begin{figure}[t!]
    \centering
    \scalebox{0.8}{
    \begin{tikzpicture}
       \tikzset{vertex/.style = {shape=circle,draw,minimum size=2em}}
       \tikzset{vertex2/.style = {shape=circle,draw,minimum size=2.5em}}
       \tikzset{edge1l/.style = {dashed, ->, transform canvas={xshift=-2pt, yshift=-1pt}}}
       \tikzset{edge1r/.style = {dashed, ->, transform canvas={xshift=-2pt, yshift=1pt}}}
       \tikzset{edge2l/.style = { ->, transform canvas={xshift=2pt, yshift=1pt}}}
       \tikzset{edge2r/.style = { ->, transform canvas={xshift=2pt, yshift=-1pt}}}
       
       \tikzset{edge3l/.style = {->, transform canvas={xshift=6pt, yshift=-1.5pt}, shorten >=-0.15cm}}
        \tikzset{edge3r/.style = {->, transform canvas={xshift=6pt, yshift=1.5pt}, shorten <=-0.15cm}}

       \tikzset{edge4l/.style = {->, transform canvas={xshift=-6pt, yshift=1.5pt}}}
       \tikzset{edge4r/.style = {->, transform canvas={xshift=-6pt, yshift=-1.5pt}, shorten >=-0.12cm}}
       
       \definecolor{node_red}{RGB}{255, 200, 200}
        \definecolor{node_green}{RGB}{200, 255, 200}
        \definecolor{c_green}{RGB}{0, 100, 0}
        \definecolor{c_red}{RGB}{100,0 , 0}

        \node[vertex] (a) at (0.5, 1.2) {AS 1};
        \node[vertex] (b) at (3.5, 1.2) {AS 2};
        
        \node[vertex, fill=node_red] (c) at (-0.3, 2.5) {};
        \node[vertex, fill=node_green] (d) at (1.3, 2.5) {};
        \node[vertex, fill=node_red] (e) at (2.7, 2.5) {};
        \node[vertex, fill=node_red] (f) at (4.3, 2.5) {};
        
        \node[vertex, fill=node_green] (g) at (-1, 3.5) {};
        \node[vertex, fill=node_red] (h) at (0.5, 3.5) {};
        \node[vertex, fill=node_red] (i) at (2, 3.5) {};
        \node[vertex, fill=node_green] (j) at (3.5, 3.5) {};
  
        \node[vertex, fill=node_red] (k) at (5, 3.5) {};
        
        \draw[purple] (-1.5,4.5) rectangle (-0.5,5.5) node[pos=.5] {VP 1};
        \draw[blue] (0.5,4.5) rectangle (1.5,5.5) node[pos=.5] {PB 1};
        \draw[blue] (2.5,4.5) rectangle (3.5,5.5) node[pos=.5] {PB 2};
        \draw[purple] (4.5,4.5) rectangle (5.5,5.5) node[pos=.5] {VP 2};
        
        \node (l) at (-1, 4.5){};
        \node (m) at (1, 4.5){};
        \node (n) at (3, 4.5){};
        \node (o) at (5, 4.5){};

        \draw[edge1l, c_green] (a) to (c);
        \draw[edge2l, c_red] (a) to (c);
        \draw[edge1r, c_green] (a) to (d);
        \draw[edge2r, c_red] (a) to (d);
        
        \draw[edge1l, c_red] (b) to (e);
        \draw[edge2l, c_green] (b) to (e);
        \draw[edge1r, c_red] (b) to (f);
        \draw[edge2r, c_green] (b) to (f);
        
        \draw[edge1l, c_green] (c) to (g);
        \draw[edge2l,  c_red] (c) to (g);
        
        \draw[edge1r, c_green] (c) to (h);
        \draw[edge2r,  c_red] (c) to (h);
        
        \draw[edge1l, c_green] (d) to (h);
        
        \draw[edge1l, c_red] (e) to (h);
        \draw[edge2l, c_green] (e) to (h);
        
        \draw[edge1l, c_red] (e) to (i);
        \draw[edge2l, c_green] (e) to (i);
        
        \draw[edge1l, c_red] (f) to (j);
        \draw[edge2l, c_green] (f) to (j);
        
        \draw[->, c_green] (j) to (k);

        \draw[->, c_green] (j) to (i);

        \draw[dashed, ->, c_green] (d) to (e);

        \node[rotate=90] at (-2.2, 3.0 ){Internet};
        \draw[dotted] (-1.8, 1.9) rectangle (5.8, 4.1);
        
        \draw[edge1l, c_red] (f) to (j);
        \draw[edge2l, c_green] (f) to (j);
        
        \draw[->, c_green] (g) to (l);
        
        \draw[edge1l, c_red] (h) to (l);
        \draw[edge2l, c_red] (h) to (l);
        
        \draw[edge1r, c_red] (h) to (m);
        \draw[edge2r, c_red] (h) to (m);
        
        \draw[edge1r, c_red] (i) to (n);
        \draw[edge2r, c_green] (i) to (n);
        
        \draw[->, c_green] (k) to (o);
        
        \draw[edge3l] (m) to (h);
        \draw[edge3l] (h) to (c);
        \draw[edge3r] (c) to (a);

        \draw[edge4l, dashed] (m) to (h);
        \draw[edge4r, dashed, shorten <=0.12cm] (h) to (e);
        \draw[edge4r, dashed] (e) to (b);

        \draw[edge3l] (n) to (i);
        \draw[edge3r] (i) to (e);
        \draw[edge3r] (e) to (b);
        
        \draw[edge4l, dashed, shorten >=0.04cm] (n) to (i);
        
        \draw[edge4r, dashed] (i) to (e);

        \draw[->, very thick, c_green, dashed] (-2.3,0.2) --++ (0.5,0) node[right]{Valid Announcement p1};
        \draw[->, very thick, c_red, dashed] (2.0,0.2) --++ (0.5,0) node[right]{Invalid Announcement p1};
        
        \draw[->, very thick, c_green] (-2.3,-0.2) --++ (0.5,0) node[right]{Valid Announcement p2};
        \draw[->, very thick, c_red] (2.0,-0.2) --++ (0.5,0) node[right]{Invalid Announcement p2};
        
        \draw[->, very thick] (-0,-0.6) --++ (0.5,0) node[right]{Traceroute Packet};
            \end{tikzpicture}
    }
    \caption[Overview RPKI Process]%
    {Measurement setup with AS 1 and AS 2 with prefixes p1 and p2. Vantage Points 1 and 2 collect received announcements, the probes PB1 and PB2 use received paths to send out Traceroutes to both prefixes.}
    \label{fig:setup}
    \vspace{-10pt}
    
\end{figure} 

Similarly to Cloudflare, the Asia-Pacific Network Information Centre (APNIC) runs an experiment to test ROV deployment over the reachability of destinations with varying ROV validity \footnote{\url{https://stats.labs.apnic.net/rpki}}. 
The measurement probes how many users in a specific AS can reach an invalid prefix to draw conclusions on the ROV enforcement status of that user’s AS. 
The APNIC measurement improves over the Cloudflare project in two keys aspect. First, they do not rely on users to visit the website and then manually contribute the enforcement status of their provider over a Github page. Instead, they automatically execute the measurement on client systems. Second, they use anycast to inject their route from a large amount of routers instead of two points used by Cloudflare. This significantly reduces the rate of false-positives. ROV enforcement in intermediate systems is less likely to affect the measurement if the route is propagated over many different paths to a target. In Section \ref{sc:rov} we use APNIC's dataset to validate our measurements and provide insights into the limitations and challenges inherent in comparing different ROV measurement methodologies. 
\section{Methodology}\label{sc:method}

The idea behind our measurements is to evaluate the route that packets traverse to a destination prefix in two scenarios: when the target prefix is hijacked and conflicts with the origin AS in the ROA covering that prefix, and when that prefix is legitimate and matches the ROA. ASes that fall victim to the hijack are classified as non-enforcing, while ASes that adapt their routing according to ROV validation results are classified as enforcing ROV.

To run ethical experiments, we hijack the prefixes on self-managed/owned resources and probe the reaction of ASes both on the control-plane and on the data-plane. We minimize any interference between Internet Routing Registry (IRR) based filtering and our experiment by creating and maintaining proper IRR records (route, route6, aut-num, etc. objects) for all announcements. This ensures that IRR-based filters allow all prefix-origin pairs announced by us to the Default Free Zone.

Our study focuses on ROV enforcement in IPv4, as it is still the predominant technology in today's Internet and differences in routing between IPv4 and IPv6 are beyond the scope of this work. However, the presented methodology is directly applicable to IPv6 measurements.

\subsection{Methodology Derivation}
We explain our methodology by first introducing approaches of the most relevant previous work: Hlavacek et al \cite{hlavacek2018practical} and Rodday et al \cite{rodday2021revisiting}. We illustrate their core ideas and identify their shortcomings. We then derive our methodology by using strengths of the presented approaches while improving on identified shortcomings with new techniques.
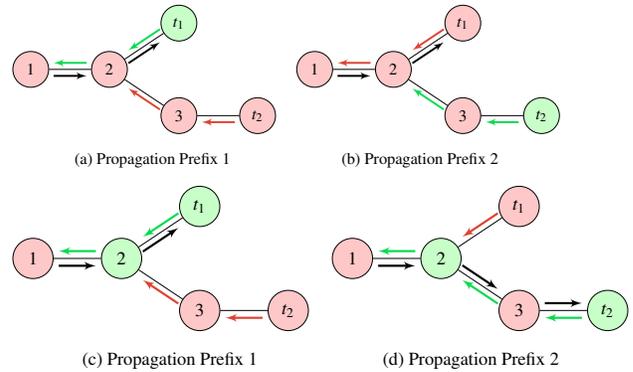
\begin{figure}[t!]
\resizebox{0.4\textwidth}{!}{%
\begin{minipage}[t]{0.34\textwidth}
        \begin{tikzpicture}
            \definecolor{c_red}{RGB}{220, 68, 51}
            \definecolor{c_green}{RGB}{11, 218, 81}
            \definecolor{node_red}{RGB}{255, 200, 200}
            \definecolor{node_green}{RGB}{200, 255, 200}

            \tikzset{vertex/.style = {shape=circle,draw,minimum size=2.2em}}
            \tikzset{edge/.style = {->, very thick, shorten >=0.1cm,shorten <=0.1cm, green, transform canvas={yshift=4pt}, > = latex'}}
                    \tikzset{edge_f/.style = {->, very thick, shorten >=0.1cm,shorten <=0.1cm, transform canvas={yshift=-4pt}, > = latex'}}
        
            % vertices
            \node[vertex, fill=node_red] (b) at  (1.7,0) {1};
            \node[vertex, fill=node_red] (c) at  (3.4,0) {2};
            \node[vertex, fill=node_green] (d) at  (4.9,1) {$t_1$};
            \node[vertex, fill=node_red] (g) at  (4.9,-1) {3};
            \node[vertex, fill=node_red] (f) at (6.6,-1) {$t_2$};
            
            %edges
            \draw (b) to (c);
            \draw (c) to (d);
            \draw (c) to (g);
            \draw (g) to (f);
            
            \draw[edge_f] (b) to (c);
            \draw[edge_f] (c) to (d);
        
            \draw[edge, c_green] (c) to (b);
            \draw[edge, c_green] (d) to (c);
            \draw[edge, c_red, transform canvas={yshift=-8pt}] (f) to (g);
            \draw[edge, c_red, transform canvas={yshift=-8pt}] (g) to (c);

            \end{tikzpicture}
            \captionsetup{labelformat=empty}
            \captionof*{figure}{(a) Propagation Prefix 1}
     \end{minipage}\vspace{50mm}
     \begin{minipage}[t]{0.3\textwidth}
        \begin{tikzpicture}
            \definecolor{c_red}{RGB}{220, 68, 51}
            \definecolor{c_green}{RGB}{11, 218, 81}
            \definecolor{node_red}{RGB}{255, 200, 200}
            \definecolor{node_green}{RGB}{200, 255, 200}

            \tikzset{vertex/.style = {shape=circle,draw,minimum size=2.2em}}
            \tikzset{edge/.style = {->, very thick, shorten >=0.1cm,shorten <=0.1cm, green, transform canvas={yshift=4pt}, > = latex'}}
                    \tikzset{edge_f/.style = {->, very thick, shorten >=0.1cm,shorten <=0.1cm, transform canvas={yshift=-4pt}, > = latex'}}
        
            % vertices
            \node[vertex, fill=node_red] (b) at  (1.7,0) {1};
            \node[vertex, fill=node_red] (c) at  (3.4,0) {2};
            \node[vertex, fill=node_red] (d) at  (4.9,1) {$t_1$};
            \node[vertex, fill=node_red] (g) at  (4.9,-1) {3};
            \node[vertex, fill=node_green] (f) at (6.6,-1) {$t_2$};
            
            %edges
            \draw (b) to (c);
            \draw (c) to (d);
            \draw (c) to (g);
            \draw (g) to (f);
            
            \draw[edge_f] (b) to (c);
            \draw[edge_f] (c) to (d);
        
            \draw[edge, c_red] (c) to (b);
            \draw[edge, c_red] (d) to (c);
            \draw[edge, c_green, transform canvas={yshift=-8pt}] (f) to (g);
            \draw[edge, c_green, transform canvas={yshift=-8pt}] (g) to (c);

            \end{tikzpicture}
            \captionsetup{labelformat=empty}
            \captionof*{figure}{(b) Propagation Prefix 2}
            \hspace{60mm}
     \end{minipage}
     }

     \resizebox{0.45\textwidth}{!}{%
\begin{minipage}[t]{0.34\textwidth}
        \begin{tikzpicture}
            \definecolor{c_red}{RGB}{220, 68, 51}
            \definecolor{c_green}{RGB}{11, 218, 81}
            \definecolor{node_red}{RGB}{255, 200, 200}
            \definecolor{node_green}{RGB}{200, 255, 200}

            \tikzset{vertex/.style = {shape=circle,draw,minimum size=2.2em}}
            \tikzset{edge/.style = {->, very thick, shorten >=0.1cm,shorten <=0.1cm, green, transform canvas={yshift=4pt}, > = latex'}}
                    \tikzset{edge_f/.style = {->, very thick, shorten >=0.1cm,shorten <=0.1cm, transform canvas={yshift=-4pt}, > = latex'}}
        
            % vertices
            \node[vertex, fill=node_red] (b) at  (1.7,0) {1};
            \node[vertex, fill=node_green] (c) at  (3.4,0) {2};
            \node[vertex, fill=node_green] (d) at  (4.9,1) {$t_1$};
            \node[vertex, fill=node_red] (g) at  (4.9,-1) {3};
            \node[vertex, fill=node_red] (f) at (6.6,-1) {$t_2$};
            
            %edges
            \draw (b) to (c);
            \draw (c) to (d);
            \draw (c) to (g);
            \draw (g) to (f);
            
            \draw[edge_f] (b) to (c);
            \draw[edge_f] (c) to (d);
        
            \draw[edge, c_green] (c) to (b);
            \draw[edge, c_green] (d) to (c);
            \draw[edge, c_red, transform canvas={yshift=-8pt}] (f) to (g);
            \draw[edge, c_red, transform canvas={yshift=-8pt}] (g) to (c);

            \end{tikzpicture}
            \captionsetup{labelformat=empty}
            \captionof*{figure}{(c) Propagation Prefix 1}
     \end{minipage}\vspace{50mm}
     \begin{minipage}[t]{0.3\textwidth}
        \begin{tikzpicture}
            \definecolor{c_red}{RGB}{220, 68, 51}
            \definecolor{c_green}{RGB}{11, 218, 81}
            \definecolor{node_red}{RGB}{255, 200, 200}
            \definecolor{node_green}{RGB}{200, 255, 200}

            \tikzset{vertex/.style = {shape=circle,draw,minimum size=2.2em}}
            \tikzset{edge/.style = {->, very thick, shorten >=0.1cm,shorten <=0.1cm, green, transform canvas={yshift=4pt}, > = latex'}}
                    \tikzset{edge_f/.style = {->, very thick, shorten >=0.1cm,shorten <=0.1cm, transform canvas={yshift=-4pt}, > = latex'}}
        
            % vertices
            \node[vertex, fill=node_red] (b) at  (1.7,0) {1};
            \node[vertex, fill=node_green] (c) at  (3.4,0) {2};
            \node[vertex, fill=node_red] (d) at  (4.9,1) {$t_1$};
            \node[vertex, fill=node_red] (g) at  (4.9,-1) {3};
            \node[vertex, fill=node_green] (f) at (6.6,-1) {$t_2$};
            
            %edges
            \draw (b) to (c);
            \draw (c) to (d);
            \draw (c) to (g);
            \draw (g) to (f);
            
            \draw[edge_f] (b) to (c);
            \draw[edge_f, transform canvas={yshift=8pt}] (c) to (g);
            \draw[edge_f, transform canvas={yshift=8pt}] (g) to (f);
        
            \draw[edge, c_green] (c) to (b);
            \draw[edge, c_red] (d) to (c);
            \draw[edge, c_green, transform canvas={yshift=-8pt}] (f) to (g);
            \draw[edge, c_green, transform canvas={yshift=-8pt}] (g) to (c);

            \end{tikzpicture}
            \captionsetup{labelformat=empty}
            \captionof*{figure}{(d) Propagation Prefix 2}
     \end{minipage}
     }
     \vspace{6pt}
     \caption{ Propagation of BGP routes. Green marks valid and enforcing components, red marks invalid and non enforcing components, black arrows indicate data-plane paths of traffic.}
     \label{fig:update_prop}
\end{figure}     

{\bf Hlavacek et al \cite{hlavacek2018practical}}.
The approach of Hlavacek et al. first introduced the concept of measuring ROV enforcement on the data-plane additionally to control-plane measurements. For the data-plane measurements, RIPE Atlas is used, a collection of small devices distributed in different ASes of the Internet. Researchers can obtain access to those devices to run traceroute measurements from many observation points to a predetermined target. The traceroutes allow the reconstruction of AS paths that BGP routes from a specific AS take through the Internet. The process is illustrated in Figure \ref{fig:setup}.
Two different origin ASes, AS 1 and AS 2, both announce the same two prefixes p1 and p2. Additionally, ROA objects are created that make the announcement of one prefix valid from the first AS, and the announcement of the second prefix valid for the second AS resulting in two valid announcements (p1-AS1 and p2-AS2) and two attempted hijacks (p1-AS2 and p2-AS1). The routing paths of valid and invalid announcements are then compared. If any traceroute to a prefix is routed to a ROA-invalid AS, i.e., falls victim to the hijack, the path to that AS is considered invalid and all ASes on the path are marked as not enforcing ROV.

We visualize the core idea of how ROV enforcement is measured with the setup of Hlavacek et al. in Figure \ref{fig:update_prop}. Subfigures (a) and (b) show the propagation of the updates announced by the two target ASes $t_1$ and $t_2$ in a scenario where no on-path AS enforces ROV. In case (a) the announcement of $t_1$ is valid, while in case (b) the announcement of $t_2$ is valid. The valid origin AS of an announcement and ROV-enforcing ASes are marked in green, while red indicates invalid announcements, propagation, and non-enforcement. 

In (a) no AS enforces ROV and thus other routing mechanisms influence the propagation of updates. In this case, AS 2 prefers the valid announcement over the invalid announcement because it has a shorter AS path to the target (1 hop to $t_1$ vs. 2 hops to $t_2$), by chance following ROA validity. For the second prefix in (b) preferring the shorter AS path conflicts with the ROA; ASes 1 and 2 thus fall victim to the prefix hijack of $t_1$. ASes in this scenario are correctly classified as not enforcing ROV.

Figures (c) and (d) illustrate a scenario where one on-path AS enforces ROV (AS 2). In this scenario both prefixes (c) and (d) are routed to the correct target; the prefix hijack is unsuccessful. AS 2 discards the hijack of $t_2$ in (c) and $t_1$ in (d). In this configuration the classification scheme of \cite{hlavacek2018practical} would classify ASes 1, 2, and 3 as ROV-enforcing, as they all do not fall victim to the hijack.

This example illustrates a shortcoming in the methodology; the classification is susceptible to false positives. In this example, AS 1 and AS 3 are wrongfully classified as ROV-enforcing. While the false positives might be reduced by using multiple origins, a lack of identification which on-path AS enforces ROV still leads to faulty classifications. Further, the methodology does not distinguish between ASes that use ROV in a non-strict mode, and ASes that do not enforce ROV.

{\bf Rodday et al \cite{rodday2021revisiting}.}
Improving on previous work, Rodday et al. develop a methodology that emphasizes mitigating false positives in their results. They use a single ASN to announce updates to the Internet and, similarly to \cite{hlavacek2018practical}, probe the paths that updates take over a large number of distributed RIPE Atlas probes. However, the methodology does not simply look at the number of valid paths over different ASes. Instead, they apply a strict classification scheme that limits false positives. They distinguish between ASes one hop away from their target and ASes 2+ hops away. ASes in a distance of one hop do not, by definition, have any AS between them and are thus not susceptible to false positives induced by other on-path ASes enforcing ROV. In the 2+ hop case, intermediate ASes may enforce ROV. The methodology thus proposes strict rules that prevent a false positive from ROV-enforcing ASes on the path. The introduced rules require that every on-path AS hosts a measurement probe to conclude enforcement, and that no other AS on the path enforces ROV. Mandating a probe in every on-path AS allows the classification to pinpoint which AS enforced ROV and which AS is only passively protected.

While this methodology likely achieves the goal of reducing false positives, it trades the reduction in false positive with an increase in false negatives. Consider again the examples in Figure \ref{fig:update_prop}. In the example (a,b) in Figure \ref{fig:update_prop}, the methodology of \cite{rodday2021revisiting} would correctly assert that ASes 1, 2 and 3 do not enforce ROV, as no strict rules are applied for non-enforcement. However, the second scenario (c,d) in Figure \ref{fig:update_prop} illustrates the limitation of the methodology regarding false negatives. The methodology requires all on-path ASes to host a probe to conclude ROV enforcement. Consider that in the scenario (c,d) in Figure \ref{fig:update_prop}, AS 1 does not host an Atlas probe. In this case, the paths observed in (c) and (d) need to be discarded, not counting the ROV enforcement in AS 2. On the other hand, the paths of (a) and (b) would be counted towards non-enforcement. Due to the unbalanced burden of proof between enforcement and non-enforcement, the methodology favors counting paths that only contain non-enforcing ASes. In contrast, paths with ROV-enforcing ASes often need to be discarded. In the presented scenario, the methodology would conclude 0\% enforcement despite actual enforcement of 33\%.

This tendency of many false negatives is also indicated in the results presented in \cite{rodday2021revisiting}; in the set of ASes that are one hop away from the observation point, 82\% of ASes exhibit some signs of direct or indirect ROV enforcement. This rate drops to 1.6\% for ASes 2+ hops away, indicating that the methodology favors the classification of non-enforcing ASes over enforcing ASes in the case of 2+ hops of distance.

Our methodology uses the insights gained by \cite{hlavacek2018practical} and \cite{rodday2021revisiting}. We use the approach of \cite{hlavacek2018practical}, i.e., announcing two prefixes from two ASes, as the basis for our measurement. However, we extend the methodology differently than \cite{rodday2021revisiting} to reduce false positives. Instead of strict rules, we pinpoint which AS actually enforces ROV and which AS is only passively protected with a new metric that we developed, referred to as divergence points.

A divergence point refers to the AS where the path to the two prefixes diverges, following the ROA validity of the prefix announcements. Again consider the example in Figure \ref{fig:update_prop}. In this case, the path to prefix 1 (c) and 2 (d) is identical in the first hop AS 1. The paths diverge after the second hop, and AS 2 would thus be considered the divergence point of paths. AS 2 is the most likely point of ROV enforcement. Our methodology classifies ASes repeatedly appearing as divergence points in different configurations as likely ROV-enforcing. In contrast, ASes that only appear on valid paths but lack identification as divergence points are considered either upstream protected or lacking sufficient evidence for either enforcement or non-enforcement. In the presented example, AS 1 would be classified as upstream protected, AS 2 as enforcing ROV, and AS 3 as lacking sufficient evidence for either classification.

\subsection{Data Acquisition}
{\bf Control-plane.} On the control plane, we use valid and invalid (hijacked) BGP announcement propagation as a metric to identify ASes with ROV enforcement. The hijack is executed from two ASes with two neighboring prefixes, as shown in Figure \ref{fig:setup}. Each AS announces the same two prefixes p1 and p2. We also create two ROAs that authorize prefix p1 from AS 1 and p2 from AS 2. Therefore, when AS 1 announces p2 (resp. AS 2 announces p1), it effectively appears as an attempted hijack of p2 by AS 1 (resp. p1 by AS 2). ASes that do not react to the hijack and route traffic for p1 to AS 2 or traffic for p2 to AS 1, despite the conflicting ROA, are classified by us as non-ROV-enforcing ASes. The robustness of the measurement is enhanced by changing the configuration between the experiments, i.e., authorizing p1 for AS 2 and p2 for AS 1 with corresponding ROAs, reducing the noise from routing events unrelated to ROV. We apply the classification scheme of \cite{hlavacek2018practical} to our results.

{\bf Data-plane.} To find which ASes route traffic for p1 to AS 2 and for p2 to AS 1, we send out traceroute probes to our target ASes. We process the resulting paths to analyze which ASes enforce ROV and which fall victim to the hijacks. 

\subsection{Data Analysis}\label{sc:classification}
{\bf Divergence points.} Our methodology provides new key aspects which result in a more accurate approximation of real-world ROV enforcement on the Internet. Our classification of the data-plane measurements incorporates information about path divergence points, which was not considered in previous studies. 
Divergence points indicate that an AS reached a different conclusion for route propagation between the two prefixes, which provides strong evidence on ROV enforcement. This additional metric improves the accuracy of the classification as it approximates the location of the ROV-enforcing AS. To remove false positives caused by ROV in an invisible IXP in front of an AS, we additionally map collected IP addresses to the routing LANs of IXPs.

{\bf AS classification scheme}. We use information about divergence points and path structure to derive a more fine-grained classification scheme that includes conclusions about invalid-route depreferencing. ASes that show evidence of divergence points but are also traversed by invalid paths likely apply ROV, but the implementation is either non-strict (AS drops invalid announcements in certain scenarios), or the decision process depreferences invalid routes but propagates them if no other routes are available. 
Further, we use the relative position of ASes on paths to conclude about the passive upstream protection of ASes without their own enforcement. The categories are defined next.

$\bullet$ \textit{Non-enforcing C1:}
These are all ASes on invalid paths, i.e. ASes that fall victim to hijacks, which indicates that they are not enforcing ROV. The ASes in this category do not have any hints for partial ROV deployment or invalid-route depreference.
    
$\bullet$ \textit{Weak depreference C2:} ASes that show some sign of ROV enforcement but were on at least one invalid path. ROV enforcement is indicated by twice as many valid than invalid paths and at least one divergence point.
    
$\bullet$ \textit{Strong depreference C3:} ASes that are most likely enforcing ROV since they have at least three times more valid than invalid paths and are a divergence point at least once in each configuration, but their enforcement is non-strict.

$\bullet$ \textit{No negative evidence C4:} All ASes that do not conflict with the ROA, but also do not show any positive evidence for ROV enforcement. It is thus unclear if they enforce ROV or are protected by ROV in other ASes.    

$\bullet$ \textit{Passive positive evidence C5:} Similar to category 4 but with an additional requirement that the paths over an AS indicate that all upstreams enforce ROV. The protection is present in all ASes that appear behind enforcing ASes.  

$\bullet$ \textit{Direct positive evidence C6:} These ASes show signs of ROV enforcement, but the evidence is not comprehensive. The AS has been on at least one path in each configuration and a divergence point at least once.   

$\bullet$ \textit{Strong positive evidence C7:} ASes with strong evidence that they enforce ROV. The AS was on a path to each prefix in each configuration and on a divergence point at least once in each configuration.

\subsection{Correlation Control- and Data-Plane}
Our methodology uses additional control-plane measurements to allow for validation of data-plane results. For this validation, we look at the overlap between the two measurements, first in the location of the vantage points and then in the overlap in classification. We discuss the overlap of measurement locations in Section \ref{subsec:cd}. We expect that both approaches should lead to a similar result on the enforcement status for each AS in the intersection set. This
hypothesis is validated using a similarity measure that correlates
control-plane categories and data-plane categories according
to their logical similarity. The mapping uses the control-plane
category as the first integer in the tuple and the data-plane
category as the second integer. This allows to calculate similarity for each AS in the intersection. For example, an AS that
is categorized into category 1 in the control-plane and category 1 in the data-plane, which results in the correlation tuple (1,1)
and that AS is thus rated as having a high similarity between
measurements. In contrast, an AS that is rated as ROV-enforcing in the control-plane as category 3 and non-enforcing in
the data-plane as category 1 results in a tuple (3,1) which is
mapped to a low similarity between results for this AS.

{\scriptsize
\begin{verbatim}
// High similarity
H = {(1,1),(1,2),(2,3),(2,4),(2,5),(3,6),(3,7),(4,5),(4,6),(4,7)}
// Medium similarity
M = {(1,3),(2,6),(2,7),(3,3),(3,4),(3,5),(4,3),(4,4)}
// Low similarity
L = {(1,4),(1,5),(1,6),(1,7),(2,1),(2,2),(3,1),(3,2),(4,1),(4,2)}
\end{verbatim}
}

High similarity refers to ASes that are mapped identical
or almost identical in both approaches, e.g., both ASes are classified as strictly ROV-enforcing. A medium similarity
between ASes does not require identical results, but the classification must still be coherent, i.e., both classification can
result from the same behavior. For example, consider an AS that allows invalid routes in certain scenarios, such as in cases when the announcement comes from a child. Then the AS would be classified as non-enforcing in the control-plane, as
it allowed an invalid route to pass. On the other hand, the data-plane has the additional measure of divergence points, it can
thus identify that the AS generally enforces ROV. Therefore the AS will be classified as strongly depreferencing invalid
routes (category-3). The classification of control-plane and data-plane is not identical and would thus refer to this result as
medium-similarity. The classification result is still coherent.

\section{Measurements of ROV-Enforcement}\label{sc:measurements}
In this section we explain the setup and experiments.
\begin{figure}[t!]
    \centering
    \includegraphics[width=0.42\textwidth]{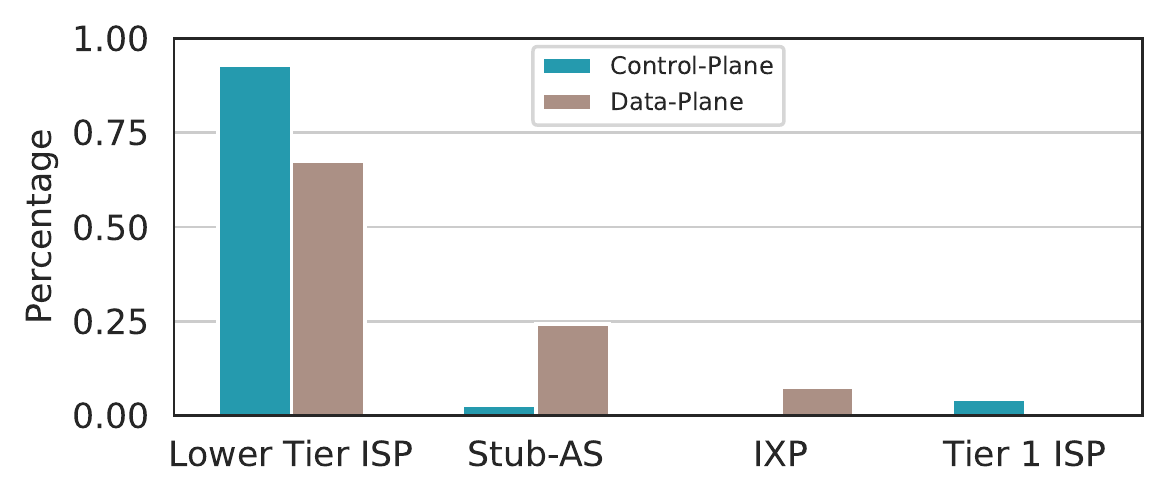}%width=\columnwidth
    \caption[Type Distribution in Data-Plane]%
    {Observed AS types in control-plane and data-plane. The relative percentage of Tier-1 ISPs is lower for the data-plane because of a substantially higher total AS amount.}
    \label{fig:as_types}
\end{figure}
\subsection{Control-Plane} On the control-plane, we carry out prefix hijacks of prefixes we own, and we use public collectors to monitor the propagation of the valid and invalid BGP announcements in the Internet.

{\bf Setup.} For our control plane measurements, we set up three origin servers, two servers by the Internet provider IBM, located in Sao Paolo (Brazil) and Tokyo (Japan), and one in a scientific institution in Germany. Both servers by IBM are assigned the AS number 212795, and the research institution server receives the AS number 208162. We use these servers to issue alternating valid and invalid BGP announcements. We also create corresponding ROAs, some valid and some conflicting with the BGP announcements. The ROAs are published in our RPKI repository.  

{\bf Monitoring.} To monitor the propagation of our BGP announcements on the control-plane, we use data from route collectors by Routeviews \cite{routeviews} and the RIPE Routing Information Service (RIS) \cite{RIS}. The collectors are BGP-speaking routers that aggregate BGP messages from peers at their respective locations and publish the collected data on the Internet. We download the data from these Vantage Points (VPs) in the form of Multi-Threaded Routing Toolkit (MRT) BGP Table dumps during the measurements and filter it for paths that originate in one of our measurement ASes. 

 \begin{figure}[t!]
    \centering
    \includegraphics[width=0.42\textwidth]{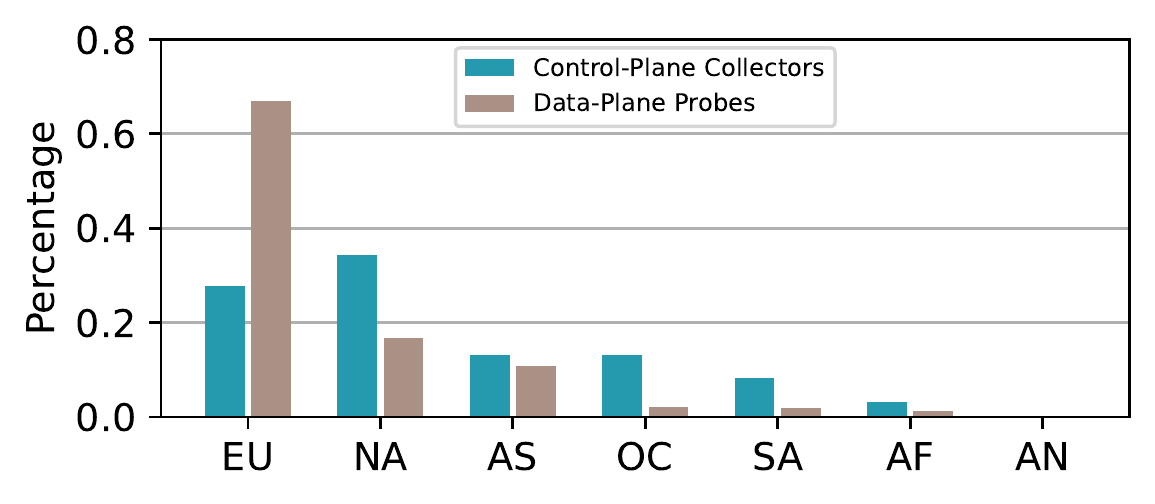}
    \caption[Probe Origin Distribution]%
    {Distribution of BGP collectors and Atlas probes over continents. The Atlas probes are biased towards Europe.}
    \vspace{-10pt}
    \label{fig:atlas_probes}
\end{figure}

%%% VISIBILITY OF THE COLLECTORS
Mapping the collectors to geo-locations, we find that the control-plane collectors are more evenly distributed over continents than the data-plane probes. We also observe that the control-plane collectors have a high presence in North America and Europe, while only a minor part is located on other continents. The distribution of the collectors is shown in comparison to the data-plane probes in Figure \ref{fig:atlas_probes}. To maximize the amount of collected data, we use all available data in both control-plane and data-plane. To understand the bias in the results from a different distribution of collectors, we present the results according to regions.
The collectors in the control-plane observe an absolute amount of 1566 paths, with 797 paths to AS212795 and 769 paths to AS208162.

{\bf Experiments.} The prefix hijacks for the ROV measurements are run on two different days, the 8th and 10th of June 2022, to increase the robustness against short-lived unexpected routing events.
All servers in our setup announce two neighboring prefixes throughout both measurements, \textit{P1} 45.155.129.0/24 and \textit{P2} 45.155.131.0/24, with their assigned AS number as the origin.

In the first configuration, a ROA is issued for AS212795 - \textit{P1} and for AS208162 - \textit{P2}. The ROAs are published 24 hours before starting the measurements. After the measurements are finished, the ROAs are withdrawn, and new ROAs are published with the inverse configuration, validating AS212795 to announce \textit{P2} and AS208162 to announce \textit{P1}. The second measurement is run six hours after the configuration change to give enough propagation time for the updated ROAs. The second measurement starts with the inverse configuration, where the ROA was published 24 hours before starting the run, preventing a one-sided bias in the results. After the measurement, the configuration is again reversed, and the final run is conducted six hours later. 

\subsection{Data-Plane} 
After executing the control-plane experiment and monitoring the propagation of the BGP announcements, we check the traffic paths on the data-plane. This analysis requires a coherent structure in the results of control-plane and data-plane. To achieve this consistency, we use Traceroute packets that probe data-plane paths through the network, resulting in a similar path structure to control-plane AS-paths. IP addresses on Traceroute paths are mapped according to the CAIDA AS and IXP mapping \cite{caidaASMapping}.

{\bf Setup.} To obtain a global distribution of origins for Traceroute measurements, we use probes by the RIPE Atlas project.

{\bf Experiments.} The data-plane measurements require a multi-step pipeline for executing the measurements, acquiring the raw data, and applying classification to the results. Measurements are started over the RIPE Atlas API. RIPE Atlas limits the measurements to 1000 probes per experiment. We start four separate Traceroute measurements per execution from 1000 random global Atlas probes, each running to both our prefixes. Experiment IDs are logged to ensure that the measurements are started from identical probes for the inverse control-plane configuration over the Atlas API.
Probes that go out of service during the measurements and thus do not complete a measurement in both configurations are removed from the results. Each measurement is run with a one-minute time difference between requests.

{\bf Processing and analysis.} We process the paths obtained from Traceroutes to remove redundant information and to discard paths that only contain unresponsive hops or originated from a probe that did not complete measurements to both announced prefixes. First, the measurement results are downloaded over the Atlas API, serialized, and receive a unique identifier that is inserted into a local database for processing. The raw data processing starts with a majority vote on each IP hop; Atlas probes run three separate Traceroutes per measurement. If no consensus between the hops can be found, the hop is added with a non-value. In the second processing step, IP addresses are mapped to AS numbers according to the CAIDA datasets \cite{caidaASMapping, caidaIXPMapping}. Traceroute logs the address of the replying interface, which may not always be correctly configured; we observed many internal IP addresses or IP addresses that could not be mapped to an AS number. These ASes are added as a non-responsive hop. Further, to prevent confusion between AS numbers and the arbitrary IDs of IXPs in the dataset, which may overlap, IXP IDs are added with a negative sign. 

The paths then need to be pre-processed to increase the robustness of classification. Consecutive hops of the same AS do not provide additional information and are thus condensed into a single hop for classification. 
None-hops are removed if the previous and following AS are identical, as the hop likely belongs to the same AS as the surrounding hops.

The data processing on the data-plane results in 18520 valid paths with 73481 valid hops and 8489 unresponsive hops. The measurements have a similar amount of paths to both our ASes, with 8286 paths to AS212795 and 7608 paths to AS208162; 2626 paths did not reach one of the target ASes.

{\bf Classification of ASes.} Applying the classification scheme from Section \ref{sc:classification} on the paths is a three-step process. First, the measurements from each probe are iterated and correlated, i.e., the paths to both prefixes in a single measurement run are processed together. Processing checks which Traceroute target follows the ROA. This process is not straightforward, as servers inside the target AS might not reply to the Traceroute. Thus we map the AS numbers of the upstream providers of our targets to the final destination. If one of the paths flows to an invalid prefix, all ASes on the path are classified as occurring on an invalid path. On the other hand, all ASes to a valid prefix receive a point of evidence for a correct path. If the paths to both prefixes follow the ROA, the divergence point between the paths is calculated, and the corresponding AS receives a classification as a divergence point. 

In the second processing step, each AS is analyzed according to its valid vs. invalid paths and divergence points. Then the classification scheme is applied. ASes are stored together with their final classification for analysis. In the last step, results for category 5 are calculated as they require information about upstream ROV enforcement. Processing iterates all ASes classified into category 4 and analyzes upstream ASes. If each path over the AS runs over an AS that is classified into category 6 or 7 before reaching one of the prefixes, the classification is changed to category 5.

\subsection{Control- vs. Data-Plane}
\label{subsec:cd}
The comparison of control-plane and data-plane measurement points plotted in Figure \ref{fig:atlas_probes} indicates that the measurements have a slightly different view of the Internet, as the control-plane has a higher percentage of points in North America while the data-plane measurements primarily originate in Europe. Additionally, the analysis of the raw data shows a different distribution of AS types between the measurements in the control and the data-plane, with a higher percentage of stub-ASes and IXPs in the data-plane, illustrated in Figure \ref{fig:as_types}.
The difference stems from the observation mode; the data-plane can observe stub-ASes because some of the stubs host an Atlas probe and are thus visible, even though they do not forward traffic. IXPs are visible because routers in their peering LAN reply to ICMP messages, even if they do not append to BGP AS paths. In contrast, IXPs and stub ASes are not visible on the control-plane. Stub ASes are not visible on the control-plane because they do not forward traffic to our destination AS. IXPs do not append their number to the control-plane path, and hence are not visible on the control plane.

\section{Results of ROV-Enforcement}\label{sc:rov}
Previous work explored the fraction of ROV filtering without characterizing the networks that enforce ROV. In 2018 \cite{hlavacek2018practical} found that between 0.5\% and 2,3\% of the ASes enforce ROV, and in 2021 \cite{rodday2021revisiting} found only 0.6\%, with more extensive protection by routeservers. Our findings indicate that at least 27\% of the ASes enforce ROV either strictly or partially, with the highest deployment rates in Europe and North America. We analyze the size and type of ROV-enforcing ASes, finding that mostly large ASes and ISPs enforce ROV.

\subsection{How many ASes enforce ROV?}

{\bf According to control-plane measurements.} The control-plane measurement passively observes BGP announcements on the Internet. Thus, it cannot directly correlate differences in forwarding paths to the two prefixes from the same origin, which mitigates the utilization of divergence points in the control-plane classification. The control-plane classification scheme thus relies on identifying ROV-enforcing ASes over negative evidence; ASes with negative evidence are classified as not ROV-enforcing. ASes without negative evidence are classified according to their visibility in different measurements. If an ASes has been observed on paths to both prefixes and in both configurations and still only forwarded valid paths, the classification scheme concludes that the AS likely enforced ROV.

\begin{figure}[t!]
    \centering
    \includegraphics[width=0.42\textwidth]{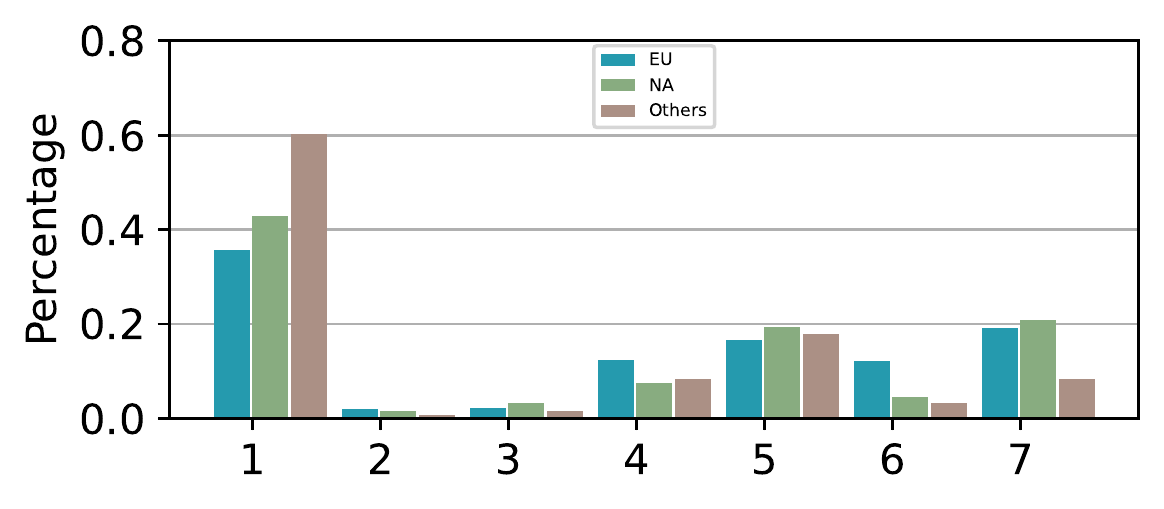}
    \caption[Classification Distribution Data-Plane]%
    {ROV enforcement worldwide. Significant differences in ROV enforcement across regions.}
    \vspace{-10pt}
    \label{fig:class-cont}
\end{figure}

The results of the control-plane illustrate an upper bound ROV enforcement of 45.4\%, summing up the observed percentages of categories 2, 3, and 4. The percentage of ASes with evidence for enforcement is 36.8\%, out of which 29.9\% show strong signs of enforcement. The distribution of ASes to control-plane categories is given by:

{\scriptsize
\begin{verbatim}
[C1] negative evidence:        190 [54.6%] 
[C2] no negative evidence:     30  [8.6%]
[C3] strong positive evidence: 104 [29.9%] 
[C4] some positive evidence:   24  [6.9%]
\end{verbatim}
}

{\bf According to data-plane measurements.} The data-plane results confirm the trend observed on the control-plane.Of the 2325 ASes observed in the data-plane measurement, approximately 24\% show signs of strict ROV enforcement. 43\% show no signs of any enforcement. This relatively low rate indicates that the majority of systems in the Internet are currently affected by ROV, either through the passive protection of ROV enforcement by others, through partial enforcement or implemented own strict enforcement. The distribution of ASes according to data-plane categories is given by:

{\scriptsize
\begin{verbatim}
[C1] no ROV:               995 [42.8%] 
[C2] weak depreference:    39  [1.7%]
[C3] strong depreference:  58  [2.5%]
[C4] no negative evidence: 393 [16.9%]
[C5] no positive evidence: 286 [12.3%]
[C6] ROV evidence:         196 [8.4%]
[C7] strong evidence:      358 [15.4%]
\end{verbatim}
}

\begin{figure}[t!]
    \centering
    \includegraphics[width=0.44\textwidth]{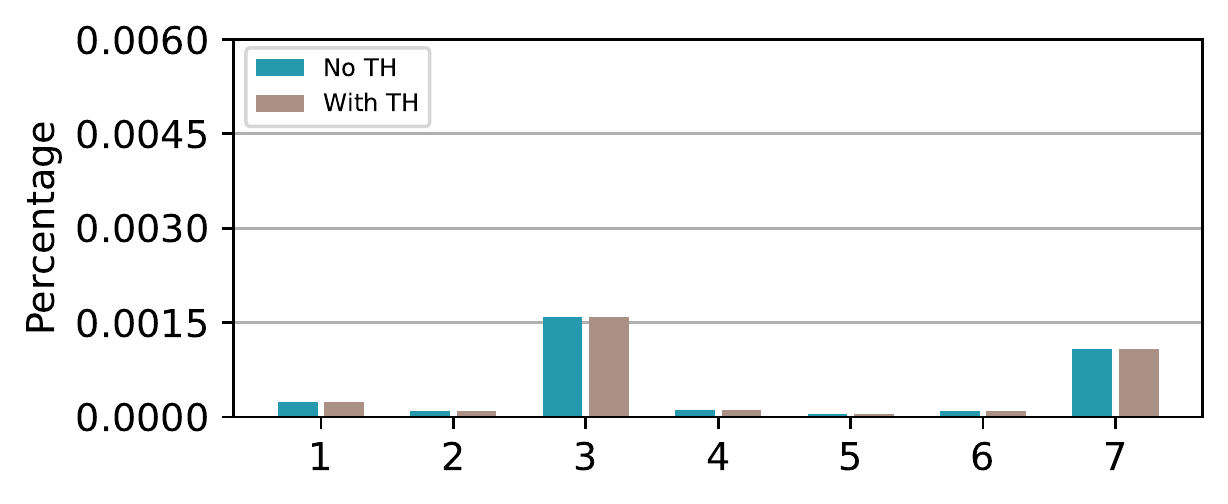}
    \caption{Size distribution among the categories. ASes with strict or partial ROV enforcement are on average larger than other categories. Allowing a threshold of invalid paths in strict enforcement increases the size in strictly enforcing ASes.}
      \vspace{-10pt}
    \label{fig:size-by-cat}
\end{figure}

{\bf Correlation control- vs. data-plane.}
Since the control-plane measurements use a different architecture than the data-plane, they see a partially different view of the Internet. Collectors of the control-plane are distributed differently than the vantage points in the data-plane and thus, our measurement results in the control-plane contain more ASes from North America. The intersection between data-plane and control-plane measurements contains 163 ASes, while 220 ASes were only observed in the control-plane, and 2162 ASes were only seen in the data-plane.

Even with its limited size, the intersection of the measurements provides insights into their classification differences.

Overall, we rate 99 ASes as high similarity, 47 ASes as medium, and 17 ASes as low similarity. These results indicate that classification results between control and data-plane are coherent; both approaches reach a consistent result for 90\% of the observed ASes. 10\% of the categorizations are conflicting, i.e., the ASes are classified as ROV-enforcing in one measurement while being classified as non-enforcing in the other. The factors that lead to a different classification include limited path visibility of the control-plane, selective enforcement depending on the route-origin, a lack of IXP visibility in the control-plane. 

\ignore{
\begin{itemize}
    \item \textbf{High Similarity} 99
    \item \textbf{Medium Similarity:} 47
    \item \textbf{Low Similarity:} 17
    \label{enum:cor-results}
\end{itemize}

\begin{enumerate}
    \item \textbf{Category 1 - No ROV:} 995 [42.8\%] 
    \item \textbf{Category 2 - Weak Depreference:} 39 [1.7\%]
    \item \textbf{Category 3 - Strong Depreference:} 58 [2.5\%]
    \item \textbf{Category 4 - No negative Evidence:} 393 [16.9\%]
    \item \textbf{Category 5 - No positive Evidence:} 286 [12.3\%]
    \item \textbf{Category 6 - ROV Evidence:} 196 [8.4\%]
    \item \textbf{Category 7 - Strong Evidence:} 358 [15.5\%]
    \label{enum:cats-dp}
\end{enumerate}
}

\ignore{
\begin{figure}[t!]
    \centering
    \includesvg[width=0.5\columnwidth]{figures_svg/categories_dp.svg}
    \caption[Classification Distribution Data-Plane]%
    {The classification results show that non-enforcing ASes are still the largest group on the Internet. However, most ASes in the measurement did not propagate ROA-invalid paths and are classified into higher categories that either indicate ROV enforcement or ROV protection by an upstream. 15.4\% of ASes have strong evidence for ROV enforcement.}
    \label{fig:categories-dp}
\end{figure}}

\subsection{Characterization of ASes with ROV}
Our measurements have good coverage of ASes and are representative. 

{\bf ROV distribution by continent.} The distribution of ROV-enforcing ASes differs significantly by continent. Figure \ref{fig:class-cont} shows that Europe and North America have significantly higher rates of enforcing ASes than the rest of the world. The higher rate of total ROV enforcement in the EU over NA (30.3\% vs. 23.5\%) might be explained by an effort of the European RIR RIPE to advance the deployment of ROV, while the North American RIR ARIN is less active in the promotion of ROV. The graph also shows that the other continents lag behind in the deployment of ROV (30.3\% vs. 11.3\%). 

\begin{figure}[t!]
    \centering
    \includegraphics[width=0.42\textwidth]{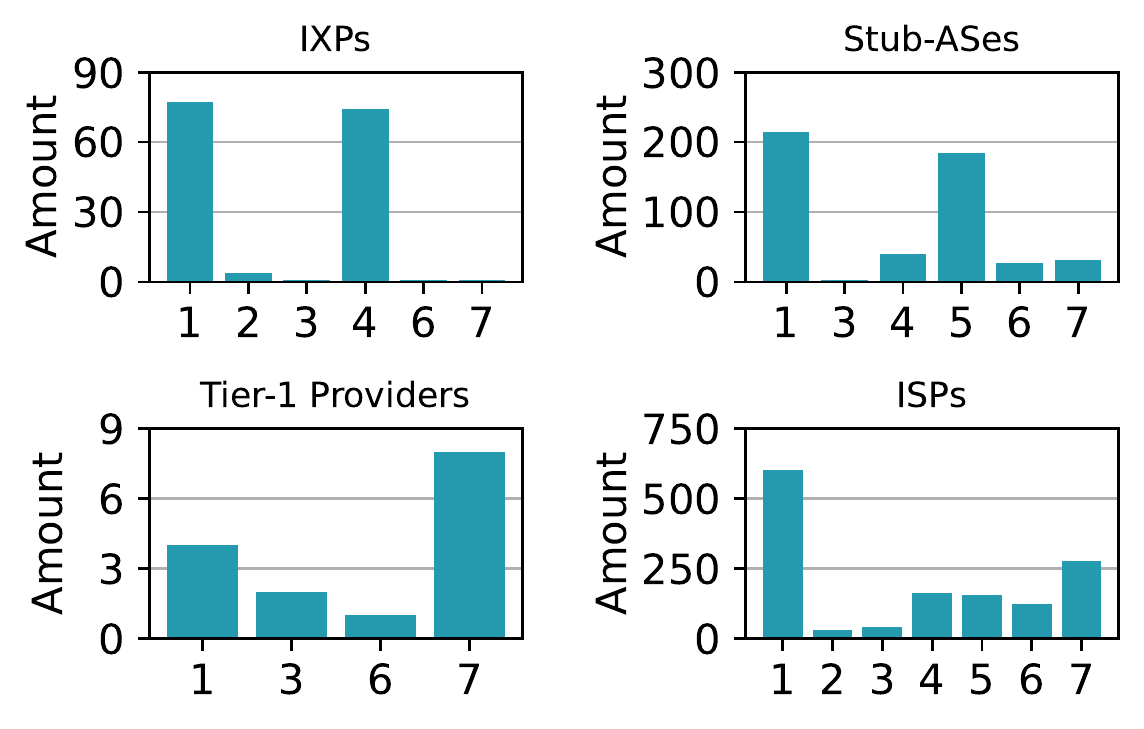}
    \vspace{-10pt}
    \caption[Classification Distribution Data-Plane]%
    {Category distribution among different AS types.}
     \vspace{-10pt}
    \label{fig:categories-types-dp}
\end{figure}

{\bf ROV distribution by AS type.} ROV protection is not equally distributed among AS types. Figure \ref{fig:categories-types-dp} illustrates that IXPs and stub-ASes have a significantly higher rate of indirect protection or the lack of positive evidence than ISPs. The measurements further indicate that nine Tier-1 providers show direct evidence for ROV enforcement, while only four providers show no signs of depreferencing invalid routes. 

{\bf AS size and ROV enforcement.}  We expect to find ASes that are classified as non-enforcing but only have a tiny percentage of invalid paths, e.g., because one router in an AS does not enforce ROV strictly. To test the impact that such ASes have on our results, we threshold the classification by the number of invalid paths as well as by the routers that forward invalid paths in an AS. We find that 10 / 1065 ASes in categories 1-3 forward less than 10\% invalid paths and have less than 10\% invalid routers, indicating either partial ROV deployment or selective route filtering. The average size of the 10 ASes, measured by the number of IP addresses in their customer cones, is 66.9x larger than the average observed AS, indicating that larger providers are more likely to apply selective filtering or have a partial deployment of ROV. 
The size distribution of ASes by category in Figure \ref{fig:size-by-cat} illustrates the difference introduced by the threshold. The threshold mostly affects large ASes, changing their classification from non-strictly enforcing to strictly enforcing. It also shows that ASes that enforce ROV strictly tend to be larger than non-enforcing ASes, even without application of the threshold.

\subsection{Accuracy of our Results}

{\bf Eliminating errors due to random events.} The room for error was minimized by routing traffic to two prefixes and using inverse configurations for the prefix hijacks. % but a small number of random routing changes might still have influenced the results.

{\bf Eliminating bias in the results.} The distribution of probes and collectors introduces a bias in the results. As the vantage points that run probes and the collectors are located in more modern parts of the Internet, the results primarily represent the technologically advanced Internet regions, e.g., Europe and North America. The smaller number of probes in Africa, Asia, and Oceania therefore limits the generalization of the results. The bias may imply that the deployment over all global ASes might be lower than the found results. To improve the visibility of the data-plane measurements, we placed two announcing servers in regions outside of Europe and North America, one in Brazil and one in Japan. 

{\bf Taking into account peering relationships.} The peering relationships between ASes influence the propagation of BGP routes. For example, a child usually does not forward routes received by its parent. Identifying these relationships is a challenging task as the relations are in constant flux and often more complex than simple child-parent or peer-to-peer peering. We consider the peering relationships between ASes implicitly. We do not construct possible paths based on assumed peering relationships between ASes. Instead, we only considers paths that we observe in our analyses. We thereby ensure that calculations and conclusions are conducted according to paths that are possible and consistent with the complex peering relationships in the real Internet.
 
{\bf Visibility of stub-ASes.} Most stub-ASes on the Internet do not run an Atlas probes and are thus not visible by the measurement. On the other hand, since most stub-ASes also do not operate a relying party validator, the ISP dataset is most relevant for measurements of ROV enforcement. 

\begin{figure}[t!]
    \centering
    \vspace{6pt}
    \includegraphics[width=0.42\textwidth]{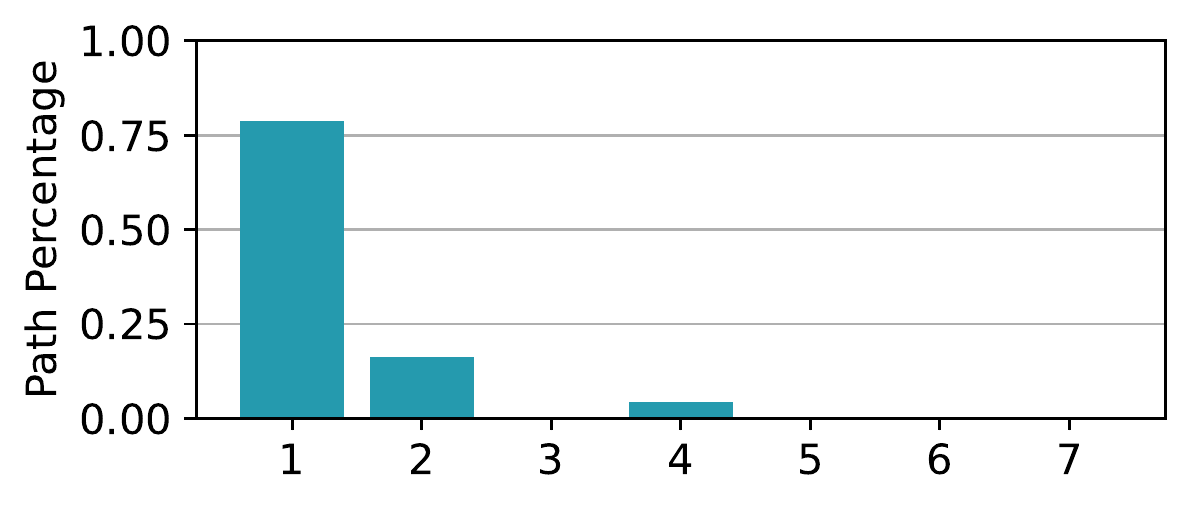}
    \vspace{-10pt}
    \caption{ Distribution of paths over IXPs by IXP category. Most observed paths run over IXPs that do not enforce ROV.}
    \vspace{-10pt}
    \label{fig:ixp-paths}
\end{figure}
 
Despite presented limitations, our methodology provides the most accurate results up-to-date, as we combine announcements from multiple ASes in multiple configurations with active measurements and a new, fine grained classification scheme. We provide a comparison of the characteristics of our study to previous approaches in Table \ref{tab:comparison}, Related Work in Section \ref{sc:works}.

\subsection{Validation of our Results}
\label{subsec:vali}
Despite ongoing interest and past discussions on the implementation of a ground-truth for ROV enforcement, no such database exists, as evident in a discussion from 3rd March 2023 on the RIPE mailing list of active experts and operators of RPKI\footnote{\url{https://www.ripe.net/ripe/mail/archives/db-wg/2023-March/007772.html}}. Therefore, validation of ROV measurements is still an open question. As shown in the linked source, the community considers the Cloudflare project as the best available data on ROV enforcement, likely as it is easy to access and work with.

We use multiple approaches to validate our results. The control-plane measurement showed that the results are mostly coherent when looking at BGP updates and data traffic. We further conduct a manual search of online sources by large providers to show that their published enforcement  status is consistent with the classification in our measurement. Additionally, we compare to two other current projects on measuring ROV enforcement, the Cloudflare project, and a project by APNIC \footnote{\url{https://stats.labs.apnic.net/rpki}}.

\textbf{Comparison to public sources.} To validate our results, we look at online publications of AS operators regarding the status of ROV enforcement in their systems. Due to the high amount of observed ASes, we limit the online search to the most significant ASes, the 15 Tier-1 providers. The search is conducted over Google with the keywords \textit{[Operator Name]}, \textit{Route Origin Validation, ROV, and RPKI}. We only consider announcements published up to 3 months after our measurement, as newer publications might indicate a change in deployment after the measurement was finished. This limitation excludes the publication of one operator, which announced ROV implementation six months after our measurements were concluded. Our search finds that all providers that we classify as either strictly enforcing ROV, or as using ROV to strongly depreference invalid routes, have made public posts announcing that they enforce ROV. We did not find any publication by the 4 providers that show no signs of ROV enforcement regarding a running deployment. One operator has published that they are working on ROV validation, but no follow-up publication has been found that announces a running deployment. Further, this operator is also classified as non-enforcing by the Cloudflare project and regularly forwards ROV invalid routes.

\textbf{Comparison to other measurements.} A comparison to other ROV measurements allows us to validate that our results are coherent with observations by other entities and that conclusions on enforcement are sensible. However, comparing to other approaches also has limitations, and a direct comparison on an AS level is only partially possible. We first compare to the Cloudflare measurement, since it uses a similar amount of route injection points, to explain discrepancies in results caused by the differences in the methodologies and limitations of the Cloudflare approach. We then present a comparison to the APNIC measurement, showing general limitations inherent in comparing both approaches with different points of route announcement. Since the APNIC approach is more sophisticated than Cloudflare, we use their results to show how different methodologies can reach differing conclusions on ROV enforcement in some systems, with both conclusions rooted in a consistent behavior of the underlying AS. This observation includes conflicting classifications of two Tier-1 providers.

{\em Comparison to Cloudflare:} Cloudflare makes their current dataset of ROV-enforcing ASes public. We compare the Cloudflare dataset to the ROV-enforcing ASes in our measurements. Our measurements observed 200 of the ASes in the Cloudflare dataset. We see an overlap in classification of 75\%, with 57.5\% being classified identically and 17.5\% classified similarly, e.g., our measurements lack positive evidence to confirm ROV enforcement while Cloudflare indicates strict enforcement or we classify it as strongly depreferencing invalid routes while Cloudflare concludes a non-enforcement. 25\% of ASes are classified differently in our results. 10.5\% are classified as ROV-enforcing in our measurements but non-enforcing in the Cloudflare set, which might indicate that the AS implemented ROV recently and no user updated the status yet, or that the AS applies selective filtering for some announcements. A prominent example of this observation is AS6461, a Tier-1 provider. Further, we see 6\% of ASes classified as ROV-enforcing, which we classify as partially enforcing. The difference is likely caused by either partial deployment, where the user that tested it was by chance protected, or by selective filtering, which again had the user protected but other traffic not protected by ROV. 5\% of ASes are classified as ROV-enforcing by Cloudflare but as completely non-enforcing by our measurements, which may indicate that the AS was upstream protected during the measurement of the user but not fully protected by all upstream providers during our measurements. The remaining 3.5\% of different classifications are attributed to changes in routing architecture as well as to errors by contributors. Thus, while there are differences in the results due to the different methodologies of our approach and Cloudflare, the results are generally coherent.

 % NEW
 
 {\em Comparison to APNIC:} APNIC also makes its measurement results available over a public API.
We thus additionally compare our findings against the results by APNIC.

To avoid conflicting classifications due to changes in the deployment over time, the results are compared on the day of our measurement. Further, since APNIC only provides results averaged out over specific time periods, we use the smallest available period of 7 days to minimize differences due to time differences between the measurements. Lastly, we exclude all ASes in the APNIC measurements that did not complete any measurement in the investigated seven days. This leaves a total overlap of 1231 ASes.

In the overlap, 971 ASes (79\%) are classified coherently in the measurements, i.e., as enforcing, non-enforcing or depreferncing in both measurements. Since the APNIC measurement averages results over seven days and a small number of invalid routes even in ROV-enforcing systems is expected, we classify an AS as enforcing in the APNIC measurement if it has more than 90\% exclusively valid measurements.

260 classifications are conflicting, including observed non-enforcement in two Tier-1 providers. Tier-1 provider AS6461 is classified as ROV-enforcing in our measurement while being classified as not enforcing in APNIC (3.16\% valid paths). Further, the AS1239 is also classified as ROV-enforcing by us while being classified as not strictly enforcing by APNIC (73.47\% valid paths). Both networks have announced publicly that they do enforce ROV \footnote{\url{https://www.sprint.net/policies/rpki}}\footnote{\url{https://seclists.org/nanog/2022/Aug/205}}.

The observed invalid routes over these networks are unlikely to originate from errors in the measurement as they are directly evident in the data. It is thus surprising that the APNIC measurements observe invalid routes despite published ROV enforcement. This observation can be explained by operational practices for implementing ROV in networks. Operators often restrain from enforcing ROV in specific BGP sessions, e.g., with their customers. For example, the online publication of AS6461 includes a statement that routes announced by the operator's customers may be excluded from ROV. While we could not find a similar public statement by AS1239, online sources by other providers indicate that exemptions from ROV enforcement in some sessions are a common practice during the implementation of RPKI \footnote{\url{https://www.gin.ntt.net/support-center/policies-procedures/routing-registry/}}\footnote{\url{https://mailman.nanog.org/pipermail/nanog/2019-February/099501.html}}\footnote{\url{https://seclists.org/nanog/2022/Aug/205}}. The need for such exemptions is also evident in RPKI documentation \footnote{\url{https://rpki.readthedocs.io/en/latest/rpki/using-rpki-data.html}}, and in the standardization of Simplified Local Internet Number Resource Management with the RPKI (SLURM) in [RFC8416]. SLURM allows administrators to override validation results of specific resources for operational purposes, for example, to allow customer routes. Thus, while the observations of the APNIC measurement conflict with the online sources on enforcement by the providers and our measurement results, they are likely caused by an APNIC anycast route injection point announcing an invalid route as a customer of AS6461 and AS1239. 

Another aspect hindering the comparison is that routing policies are usually business secrets. Thus, directly identifying if any AS, including AS6461 and AS1239, only forwarded the routes because a customer announced them is impossible with ROV measurements. However, comparing the conflicting classifications between the APNIC measurement with many injection points and our measurement with three injection points still provides insights into the different views of the Internet that different measurements observe, depending on the methodology used and the injection points.

First, we look at the overall amount of different classifications. The APNIC measurement identified 200 ASes as not enforcing ROV, which we classified as ROV-enforcing. Further, we identify 60 ASes as not enforcing ROV that are classified as enforcing by APNIC. Thus, both measurements see a substantial amount of non-enforcing ASes that are observed to enforce ROV in the other measurement. The APNIC measurement classifies more ASes non-enforcing, which appear enforcing in our measurement than vice versa.
To explain why APNIC sees more non-enforcing ASes than our measurement, we investigate ASes with conflicting classifications apart from ASes with coherent classifications.

For this, we look at the position of ASes on the Internet tree. We expect to see differences in the tree position between ASes classified conflictingly and ASes classified consistently; ASes higher in the tree have more customers and are thus more likely to have one of the route-injecting ASes as a customer. 
The position of an AS in the Internet tree is identified by calculating the minimal amount of hops of every AS to a tree root in the CAIDA Internet graph \footnote{https://snap.stanford.edu/data/as-caida.html}. For example, an AS that is a direct customer of a Tier-1 provider would be considered on the second layer of the Internet tree (1 hop). Further, the grandchild of a Tier-1 provider would be considered on the third layer (2 hops).

On average, we find that ASes observed in both measurements are 1.42 hops away from a Tier-1 provider, i.e., a tree root. Interestingly, both measurements see a noticeable difference between ASes that are classified as ROV-enforcing and ASes that are identified as non-enforcing. On average, ASes classified as enforcing in both measurements are 1.14 hops distant from a Tier-1 provider. This distance increases by 34\% to 1.53 hops for ASes classified as non-enforcing, indicating that ROV-enforcing ASes tend to be higher in the Internet tree than non-enforcing ASes.
The difference in relative position is a sensible observation as networks with routing as a core business focus are incentivized to make direct contracts with Tier-1 providers for better reachability, and to implement ROV to secure their routing. A correlation between the implementation of ROV and a high position in the tree is thus expected.
We further look at the tree position of ASes with conflicting classifications in the two measurements.

ASes that are classified as ROV-enforcing in our measurement but as non-enforcing by APNIC have a 20.3\% higher tree position than ASes that are classified as non-enforcing by both (1.22 vs. 1.53). Similarly, ASes that are classified as non-enforcing by our measurement but as enforcing by APNIC are 11.8\% higher in the tree (1.35 vs. 1.53).
Thus, ASes which show signs of selective route filtering are higher in the tree than ASes with consistent behavior. This is not surprising considering that selective filtering is mostly used to allow invalid customer announcements. ASes higher in the tree generally have more customers in their cone and are thus more likely to have a measurement injection point as a customer. It is thus expected that more classification conflicts are observed in ASes higher in the tree, even though these ASes have higher visibility and are thus easier to classify. 

The impact of selective filtering on the results of a measurement increases with the number of injection points. With more ASes announcing the invalid prefix, the likelihood of an announcing AS in the customer cone of a selectively filtering AS increases. Measurements with different locations and amounts of injection points see a partially different view of ROV enforcement in the Internet. The influence of the different amount of injection points is also indicated in the comparison to APNIC. The APNIC measurement has more injection points and correspondingly has stronger indications for the effects of selectively enforcing ASes than our measurement (20.3\% vs. 11.8\%). This observation indicates that adding more route injection points while reducing false-positives also increases the rate of networks observed as non-enforcing because they apply selective filtering. We conclude that there is a great benefit in applying a path-aware methodology to reduce false-positives instead of simply adding more injection points.

\section{Invalid Paths over Internet Exchanges}\label{sc:routeservers}
Previous work reports that when ASes use ROV-filtering on routeservers at Internet Exchange Points (IXPs)\footnote{An Internet Exchange is a switching platform over which customers exchange peering traffic with each other.}, they are protected from hijacks since invalid routes do not reach them \cite{reuter2018towards,rodday2021revisiting}. This conclusion was derived based on the fact that during the measurements, no invalid paths were sent from the routeservers. IXPs use routeservers to manage peering arrangements for their customer ASes present at an IXP. An AS can connect its border router with a single BGP session to the routeserver, which connects it with separate BGP sessions to the peers. In addition to the administration of peering arrangements, ROV-enforcing routeservers also guarantee protection to the customer ASes against BGP prefix hijacks by dropping invalid routes via ROV filtering at the routeserver. Indeed, the majority of large IXPs are known to enforce ROV. Therefore the IXPs promise to block invalid paths to the ASes peering over them.

In our study, we find that the data-plane paths traversed the address space of 159 different IXPs\footnote{There are ca. 300 active IXPs, of which 70 IXPs with more than 10GB/s peak throughput {\scriptsize{\url{wikipedia.org/wiki/List_of_Internet_exchange_points_by_size}}}.}, including the most significant European IXPs DE-CIX Frankfurt (on 3000 paths) and Amsterdam Internet Exchange (on 1300 paths).
Due to the prevalence of IXPs on Internet paths, we explore their role in blocking the propagation of invalid routes with ROV filtering at the routeservers. We find that the ROV enforcement at the routeservers generally does not block the global propagation of invalid routes. In this section, we explain our study and the factors that limit the security effect of routeservers. A detailed analysis of the impact of routeservers is in Section \ref{sc:invalid:paths}.

{\bf Leakage of invalid routes at IXPs.} We analyze the data-plane paths to investigate which IXPs enforce ROV. Since the number of paths over IXPs is not equally distributed, we look at the number of paths over IXPs by category in Figure \ref{fig:ixp-paths}. The graph shows that most paths run over IXPs that are classified as either non-enforcing or weakly de-preferencing invalid routes, which is surprising as large IXPs are known to enforce ROV. It would thus be expected that they are classified as strongly de-preferencing invalid routes.

{\bf Direct peerings propagate more routes.} We explain this observation by taking the five largest IXPs in our measurements as an example: DE-CIX Frankfurt, AMS-IX, Milan IX, NIX.CZ, and EQUINIX Singapore. Manual investigation of the five IXPs shows that all their routeservers implement ROV \cite{eqRPKI, aixRPKI, mixRPKI, nixRPKI, decixRPKI}, and the looking-glasses confirm that the routeservers drop invalid BGP announcements that conflict with ROAs. 

\begin{figure}[t!]
    \centering
    \includegraphics[width=0.85\columnwidth]{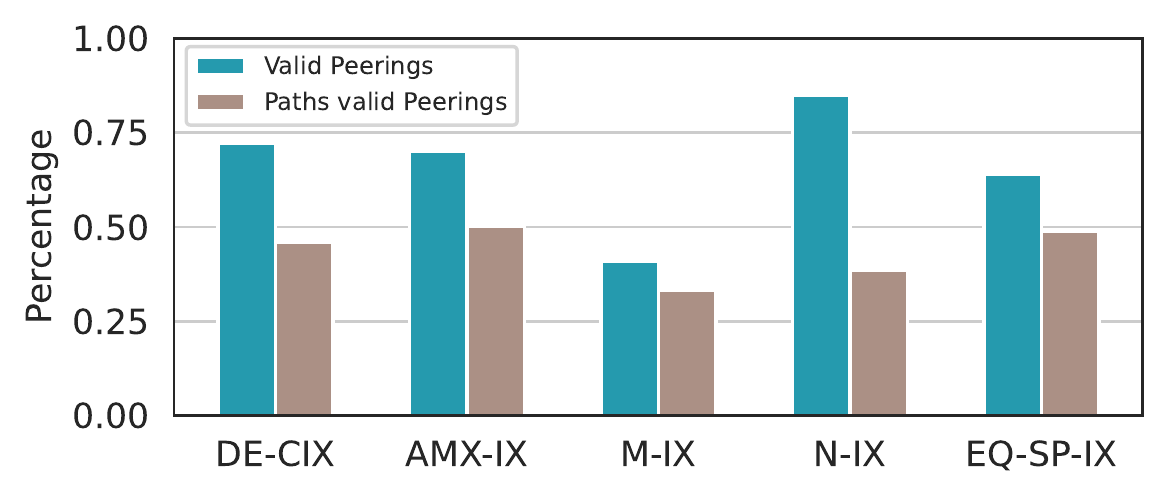}
    \vspace{-10pt}
    \caption{Percentage of connections over IXPs that only forwarded valid paths \& percentage of IXP paths that only use these valid peerings.}
    \label{fig:path-over-val-peer-paths}
\end{figure}

The observation that the routeservers of these IXPs do not propagate invalid routes leads to the conclusion that all invalid routes over their address space must have been propagated over direct sessions and not the routeserver. We thus approximate that all connections over the IXP that only propagated valid paths run over the routeserver.
Applying the approximation to the results shows that most peerings we observe in measurement paths over IXP address space, only forward valid paths and are thus considered routeserver connections. The percentage of approximated routeserver peerings is plotted in Figure \ref{fig:path-over-val-peer-paths}. While our study shows that most peerings only forward valid updates, the amount of paths in our measurement over routeserver peerings is lower than expected.
{\em We find that the non-routeserver peerings, on average, propagate more paths than the routeserver peerings}. 
Therefore, in all the top five IXPs, the total number of paths learned over direct peerings is equal to or larger than over the routeserver. The average direct peering session we observed propagates 3.4x more paths than sessions over a routeserver in the top five IXPs and 2.9x more paths averaged over all observed IXPs.

The paths learned over direct peerings are not protected by ROV in the routeserver. As a result, a significant fraction of invalid paths traverse the top five IXPs despite their ROV enforcement. We confirm our analysis with a measurement of invalid path propagation over the IXPs and find that all the five IXPs allow a significant percentage of invalid traffic over their IP space: DE-CIX Frankfurt has 33.5\% invalid paths (1000 / 2982), Amsterdam IX has 27.4\% invalid paths (342 / 1246), Milan IX has  27.9\% invalid paths (63 / 226), NIX.CZ has 32.3\% invalid paths (62 / 192), and Equinix Singapore has 40.3\% invalid paths (77 / 191).

{\bf IXPs do not block hijacks.} Our analysis shows that although routeserver ROV protects most peering connections over the IXP, it does not protect the majority of paths that traverse IXP address space. Direct peering sessions circumvent the security impact of ROV in IXPs because they allow invalid updates to leak over the IXP address space and propagate to different parts of the Internet. IXPs cannot prevent leakage of invalid paths because they do not have control-plane influence over the direct session traffic routed through their IP space. The existence of many direct peering sessions causes the ASes not to be protected since they allow the propagation of invalid announcements despite the ROV filtering at the routeserver. 

{\bf Why direct peering sessions instead of routeserver?} Many large providers create direct peering sessions or have historical relationships with other connected ASes at the IXP. In addition, the routeservers are a relatively new service, while most ASes already have existing business agreements. Finally, a significant aspect is that with the routeservers, the ASes lose control over their routing configurations and cannot have comparable fine-grained control over their path policies. For instance, the ASes need to rely on the preferences of the routserver to choose the optimal route, potentially resulting in non-optimal routing for a connected AS.
\section{Propagation of Invalid Paths}\label{sc:invalid:paths}
In this section, we consider the impact of the ROV ASes that we collected in our study on the propagation of invalid BGP announcements.
Our goal is to complement the findings in our measurements by quantifying the impact of ROV-enforcement in the observed ASes and IXPs. Graph analysis gives insights into how far the invalid paths can reach, the scope of the affected networks, and the impact of ROV on reachability. We also examine which parts of the Internet are not protected and which networks play a central role in blocking hijacks, providing global protection. To answer these questions, we develop a new graph-based analysis for measuring the propagation of invalid paths, using data-plane paths that we found in our measurements. We compare the Internet graph used by valid updates to the reduced Internet graph for invalid updates, which only includes vertices and edges without ROV-enforcement. We then analyze the differences between the graphs to derive conclusions about the impact of ROV on the propagation characteristics of invalid updates. We quantify the security of specific nodes and a general reduction of graph connectivity resulting from fewer available propagation paths. The analysis includes standard graph metrics like the number of sub-components, the node degree, the algebraic connectivity, and the average shortest- and longest-path length. 

\subsection{Graph Generation}
The graphs for the analyses are derived from the paths observed in the data-plane. The graph directly reflects the routes identified in our measurements, constituting a subset of the real-world Internet graph. 

{\bf Representing neighboring ASes.} We represent the paths and ASes as an undirected, non-cyclic graph.
Each AS on any path in the measurement is represented as a vertex in the graph, excluding IXPs. Connections between ASes that are neighbors on a path are represented as edges of two types:

\textit{Direct edges} are created from direct neighbors on a path, i.e., ASes that are topologically located in consecutive positions on the path. The edges represent a form of direct peering between the ASes, and it is expected that no intermediate party can influence the path propagation over that edge. 

\textit{Indirect edges} are edges over IXPs. These edges have one or multiple hops between the respective AS routers that belong to the peering LAN of an IXP. The ASes are neighbors in the graph because they have a peering relationship, either with a direct peering session or over a routeserver. Indirect edges differ from direct edges because they may run over a routeserver and thus be removed from the ROV graph, even if the connected ASes do not enforce ROV.

{\bf Graphs.} The resulting fully connected graph $G_1$ consists of 2156 nodes and 3810 edges. A second graph $G_2$ is created from $G_1$ to model the propagation of invalid updates by augmenting $G_1$ with information about ROV-enforcement.

In $G_2$, all edges to nodes that enforce ROV are removed from the graph as they filter out and drop invalid updates in real-world path propagation. The resulting graph is a subset of $G_1$ with the same amount of nodes but a reduced number of edges. Differential analysis of the two graphs offers insight into how much ROV impacts the graph structure and protects contained ASes. An attacker that announces a hijacked prefix can only use propagation paths in $G_2$ to reach victims , as all nodes in $G_1$ that enforce ROV would block the hijack.

An additional graph $G_3$ is created from $G_1$ to quantify the impact of ROV-enforcement in IXP routeservers. All indirect edges suspected of running over a routeserver are marked as ROV-enforcing and removed from $G_3$. The removal includes all indirect edges that only propagated valid paths in the data-plane measurement. The graph $G_3$ thus represents a scenario where ROV is only enforced in observed IXP routeservers.

\subsection{Graph Analysis}
The impact of ROV is quantified by comparing the three graphs with respect to the graph metrics. Calculating the graph metrics yields the results presented in table \ref{tab:graph}. 

\begin{table}[t!]
\renewcommand{\arraystretch}{0.6}
    \centering
    \footnotesize
\begin{tabular}{l|P{1cm}|P{1cm}|P{1cm}}
%\hline
\textbf{Graph Parameters} &  \textbf{$G_1$} & \textbf{$G_2$} & \textbf{$G_3$} \\\hline \hline
Vertices & 2156 & 2156 & 2156 \\ \hline 
Edges & 3810 & 1974 & 3173  \\ \hline
Components & 1 & 808 & 35 \\ \hline
Largest Component & 2156 & 1315 & 2110  \\ \hline
Avg. Node-Degree & 1.77 & 0.90 & 1.47  \\ \hline
Avg. Algebraic-Connectivity & 187.97 & 6.29 & 21.68  \\ \hline
Avg. Shortest-Path Length & 4.55 & 2.97 & 5.00 \\ \hline
Avg. Longest-Path Length & 9.52 & 5.78 & 9.34  \\% \hline
\vspace{-10pt}
\end{tabular}
\caption{Graph metrics for presented graphs.}
\vspace{-10pt}
\label{tab:graph}
\end{table}

\textbf{Impact ROV-enforcement on ASes.} Comparing metrics on $G_1$ and $G_2$ indicates that ROV substantially affects the measured Internet graph. ROV removed almost half of all edges for invalid updates, significantly reducing the graph's connectivity. The algebraic connectivity confirms that connectivity is decreased by more than an order of magnitude, showing a less dense mesh of connections inside the graph. ROV disconnected 808 components from the main graph. These components can be seen as isolated domains for updates; invalid messages can only spread to other parts of the component but not reach other components of the graph. The domain is also protected from any invalid updates from outside vertices. The average shortest path length between nodes in the graph is significantly reduced even though the graph is less connected, which is a direct result of the high prevalence of isolated components. The value is calculated as the average shortest path length to each reachable node from a vertex, which directly depends on the average component size. As paths inside the components are, on average, smaller than in the initial connected graph, the value reduces. 

The average longest path length, i.e., the shortest distance to the furthest distanced node in the graph for each vertex, is decreased by almost 40\%. Thus invalid updates have, on average, a 40\% shorter possible maximum AS path length than valid updates, which indicates that most invalid updates cannot propagate globally and attacks stay localized, close to the attacking AS. This reduction is in large part caused by ROV-enforcement in the Tier-1 providers. They are responsible for propagating updates over long distances across countries and continents. ROV implementation in these ASes reduces the propagation of updates from a global to a local level, as intercontinental propagation is severely limited without using Tier-1 providers. Our analysis shows that 580 edges in the graph run over an ROV-enforcing Tier-1 provider. Removing these enforcing edges is responsible for 30\% of the average longest-path reduction in $G_2$.

The graph analysis also reveals the limitations of ROV deployment on the modern Internet. ROV-enforcing ASes cannot disrupt the connectivity of the entire graph, and a significant central component of 1315 ASes remains connected in $G_2$. The remaining component can be attributed to the design principle of the Internet as a high-availability network. The Internet is a dense network of connections with a substantial amount of redundant edges, which offers robustness against outages caused by node- or edge failures. However, this design also limits the impact of ROV in single ASes. Only the removal of a large majority of ASes could result in the breakdown of the strongly-connected central component of the Internet. ASes close to the Internet's core need to implement ROV themselves for reliable protection against hijacks, as the dense mesh of connections will likely provide a propagation path for an invalid update through the Internet core, even if many ASes enforce ROV. This observation does not imply that ROV-enforcement will not significantly impact the graph. A study by Cohen et al. \cite{cohen2001breakdown} showed that removing central nodes from a scale-free network, in our case these are the Tier-1 providers for the Internet graph, can still significantly affect the connectedness and reachability of nodes in the graph. Thus ROV in central components impacts the propagation of invalid updates, even if a sizeable connected component remains on the Internet. The existence of the central components can be seen in the node degree distribution of both graphs in Table \ref{tab:graph}. 

The comparison between the graphs shows that ROV limits the impact of attacks by reducing connectivity and propagation of invalid updates on today's Internet. It localizes most attacks and hinders the global spread of hijacks by removing essential edges for global connectivity. The results indicate that ROV-enforcement in Tier-1 providers significantly impacts the spread of invalid updates as these central components play a crucial role in invalid update propagation.

\textbf{Impact ROV-enforcement on IXPs.} ROV-enforcement in IXPs does not show a similar impact to ROV in large providers. An upper limit of 637 edges in the measurement are marked as possibly running over an ROV routeserver and are removed from the graph, which is a significantly lower amount of removed edges than for $G_2$. The lower amount also reflects in the number of isolated components; only 34 components are disconnected from the main graph. The likelihood that an AS or all its upstream providers are solely connected over an ROV-enforcing routeserver appears to be too low to disrupt most parts of the graph significantly. Most ASes that we observed at IXPs or their upstream providers have direct peering sessions that leak invalid updates over the IXP, even if the AS has some or most peering connections over the routeserver. The results also show that routeserver connections provide considerable connectivity to the graph. ROV-enforcement reduces the algebraic connectivity by an order of magnitude. Invalid updates have a less dense mesh of paths available and need to take longer paths to their target, which also reflects in the increase of shortest path length in $G_3$. Longer path lengths and reduced connectivity lower the effectiveness of attacks because ASes may prefer shorter paths less and thus the hijacking announcements would lose against legitimate path announcements. Still, the protection is lower than the removal of far-reaching propagation paths. Graph $G_3$ has a minor decrease in average longest-path length; updates can propagate almost as far as in the baseline graph, even if they might have to take longer paths.

The current implementation of ROV in routeserver thus has a limited impact on the global spread of routes, while the protection they offer locally is substantial. Routeservers reduce the connectivity for invalid updates as they limit the available propagation paths to direct peering sessions. However, the prevalence of the direct sessions at today's IXPs is sufficient to allow the propagation of most invalid announcements to wider parts of the Internet. The routeservers only marginally reduce the maximum reach of hijacks, as updates leak over direct sessions and are propagated by global providers that usually do not peer at IXPs and routeservers. The effect of routeserver ROV is thus mainly localized, reducing the local connectivity for invalid updates and preventing the spread of the hijack to connected ASes that run only routeserver peering. Routeservers should thus be considered as a measure to reduce the spread of hijacks for protecting local ASes, but they cannot mitigate the global spread of hijacks in a similar capacity to Tier-1 providers.
\ignore{
\section{Future Deployment and Research}
The results raise the question of the future direction that deployment of ROV should take to reach the most efficient protection of the Internet. Counter-intuitively, routeserver ROV can not solve the problem of origin hijacks on today's Internet. The high prevalence of direct peering sessions limits the capability of IXPs to mitigate the spread of malicious updates. Further, even without the prevalence of direct peering sessions, IXPs only provide localized protection against hijacks. The burden of protecting large parts of the routing architecture lies on the large ISPs and Tier-1 providers.

ROV implementation in Tier-1 providers greatly benefits the security of the Internet as it limits the spread of hijacks to a localized scope. Thus, it is vital that the remaining seven Tier-1 providers without strict enforcement terminate the propagation of invalid updates. The implementation in Tier-1 providers should be supplemented by ROV implementation in Tier-2 ISPs and other large systems, further limiting the reach and connectivity for attacks.

Implementing ROV in more routeservers is also a crucial step towards a safer Internet. It provides the last link in the chain of protection for the Internet by extending local protection to a large amount of ASes. The combination of global protection from large providers with localized protection by IXPs can lead to a substantial amount of protected ASes without the need to implement ROV in every system.

Nevertheless, ASes should be aware that direct peering sessions at IXPs circumvent the ROV protection IXPs offer. Operators should consider moving peering sessions that do not require fine-grained control over route selection to the routeserver or implement ROV themselves. 

The results of this work motivate multiple open questions for future research.

The measurements showed that ROV implementation on routeservers does not play a central role in protecting the Internet from prefix hijacks. This result was obtained by an upper-bound estimation of paths over the routeserver, which proved that even in the best-case scenario, the protection of routeserver ROV is primarily local and does not reliably protect wider parts of the Internet. This result should be explored further in future research with measurements that can better estimate routeserver connections and their impact on invalid path propagation. 

Future research should also look into the deployment of RP clients inside ROV enforcing ASes. The measurements revealed a high number of ASes that exhibit clear signs of ROV enforcement without an RP client requesting from their IP space. Investigating this observation with further measurements might give valuable insights into the interplay of the different parts of ROV and their usage in real-world ASes. 

The measurements also found several ASes running multiple RP clients from their address space. Future research should explore the decision processes between those clients and potential changes in ROV enforcement if one or multiple clients either fail or provide conflicting validation results.

%%%%%%%%%%%%%%%%%%%%%%%%%%%%%%%%%%%%%%%%%%%%%%%%%%%%%%%%%%%%%%%%%%%%%%%%%%%%%%%%%%%%%%%%%
\section{Limitations}
Measuring real-world Internet routing always comes with limitations. As the routing infrastructure is in constant flux, a random event might influence results and distort the drawn conclusions. It is thus possible that a subset of the observed routing changes was misinterpreted as ROV-related. The room for error was minimized by routing traffic to two prefixes and using inverse configurations for the prefix hijacks, but a small number of random routing changes might still have influenced the results.

The distribution of probes and vantage points introduces a bias in the results. As the probes and vantage points are located in more modern parts of the Internet, the results primarily represent the technologically advanced Internet regions, i.e. Europe and North America. The visibility of the data-plane measurement was improved by placing two of the announcing servers in regions outside of Europe and North America, one in Brazil and one in Japan. Still, the small number of probes in Africa, Asia, and Oceania limits the generalization of the results, and we expect that the deployment over all global ASes might be significantly lower than the found results. Most stub-ASes on the Internet do not run an Atlas probe and are thus not visible by the measurement. The low absolute percentage of ASes on the Internet running an RP client indicates that the results do not allow valid conclusions about ROV enforcement in ASes other than ISPs.

The mappings of IP addresses to AS numbers and geo-locations used data-sets by the CAIDA foundation and an IP location provider. The measurements rely on the accuracy of these mappings to draw robust conclusions from the acquired data. The reliability of the data-sets is thus an additional potential error source for the measurement results.
}

\section{Conclusions}\label{sc:conclusions}
RPKI is crucial for Internet security. Not only does it block hijacks, but it also paves the foundations for other mechanisms, such as BGPsec \cite{austein2017rfc}, ASPA\footnote{\url{https://datatracker.ietf.org/doc/draft-ietf-sidrops-aspa-verification}}, RTA\footnote{\url{https://tools.ietf.org/html/draft-michaelson-rpki-rta-00}} against path manipulation attacks. These standards build on RPKI as the source of truth about the origin authorizations and routing validation. 

In this work, we develop an improved methodology for measuring ROV and find that more than 27\% of the ASes currently filter bogus BGP announcements with ROV. Our measurements are more accurate and identify more ROV-enforcing ASes than previous work. 
We show that most ROV-supporting ASes are providers who apply filtering to avoid losing their customers' traffic, indicating that ROV deployments are aligned with the BGP business incentives. Stub-ASes, which are not paid for traffic forwarding, have lower rates of ROV enforcement. This observation is also useful for other mechanisms and facilitates their deployment. 

Surprisingly, we find that ROV on routeservers cannot solve the problem of origin hijacks. Analysis of the impact of routeserves on the propagation graph shows that the high prevalence of direct peering sessions limits the capability of IXPs to mitigate the global spread of malicious paths. Further, even without the prevalence of direct peering sessions, graph analysis shows that IXPs provide only localized protection against hijacks. The burden of protecting the global routing architecture primarily lies on large ISPs and Tier-1 providers. ROV implementation in Tier-1 providers greatly benefits Internet security as it limits the spread of hijacks to a localized scope. To achieve global protection, deployment of ROV should be increased in large providers, as the current ROV is insufficient to protect all ASes and cannot reliably prevent the spread of invalid updates on a local or national scale.
A combination of global protection in large providers with localized protection in IXPs would provide optimal protection of the Internet.

\section*{Acknowledgements}
This work has been co-funded by the German Federal Ministry of Education and Research and the Hessen State Ministry for Higher Education, Research and Arts within their joint support of the National Research Center for Applied Cybersecurity ATHENE and by the Deutsche Forschungsgemeinschaft (DFG, German Research Foundation) SFB~1119.

{
\footnotesize
\bibliographystyle{IEEEtran}
\bibliography{main.bib,ref,bib,sec}

\begin{thebibliography}{10}
\providecommand{\url}[1]{#1}
\csname url@samestyle\endcsname
\providecommand{\newblock}{\relax}
\providecommand{\bibinfo}[2]{#2}
\providecommand{\BIBentrySTDinterwordspacing}{\spaceskip=0pt\relax}
\providecommand{\BIBentryALTinterwordstretchfactor}{4}
\providecommand{\BIBentryALTinterwordspacing}{\spaceskip=\fontdimen2\font plus
\BIBentryALTinterwordstretchfactor\fontdimen3\font minus
  \fontdimen4\font\relax}
\providecommand{\BIBforeignlanguage}[2]{{%
\expandafter\ifx\csname l@#1\endcsname\relax
\typeout{** WARNING: IEEEtran.bst: No hyphenation pattern has been}%
\typeout{** loaded for the language `#1'. Using the pattern for}%
\typeout{** the default language instead.}%
\else
\language=\csname l@#1\endcsname
\fi
#2}}
\providecommand{\BIBdecl}{\relax}
\BIBdecl

\bibitem{china:telecom}
Arstechnica, ``{BGP event sends European mobile traffic through China Telecom
  for 2 hours},''
  \\\url{https://arstechnica.com/informationtechnology/2019/06/bgp-mishap-sends-europeanmobile-traffic-through-china-telecom\\
  -for-2-hours}, 2019.

\bibitem{fb:out}
S.~Janardhan, ``{More details about the October 4 outage},''
  \\\url{https://engineering.fb.com/2021/10/05/networking-traffic/outage-details/},
  2021.

\bibitem{turkey:hijack}
A.~Toonk, ``{Turkey Hijacking IP Addresses for Popular Global DNSProviders},''
  \\\url{https://www.bgpmon.net/turkey-hijacking-ip-addresses-for-popular-\\
  -global-dns-providers/}, 2014.

\bibitem{bellovin1989security}
S.~M. Bellovin, ``Security problems in the tcp/ip protocol suite,'' \emph{ACM
  SIGCOMM Computer Communication Review}, vol.~19, no.~2, pp. 32--48, 1989.

\bibitem{ballani2007study}
H.~Ballani, P.~Francis, and X.~Zhang, ``{A Study of Prefix Hijacking and
  Interception in the Internet},'' in \emph{ACM SIGCOMM Computer Communication
  Review}, vol.~37.\hskip 1em plus 0.5em minus 0.4em\relax ACM, 2007, pp.
  265--276.

\bibitem{vervier2015mind}
P.-A. Vervier, O.~Thonnard, and M.~Dacier, ``{Mind Your Blocks: On the
  Stealthiness of Malicious BGP Hijacks},'' in \emph{NDSS}, 2015.

\bibitem{hlavacek2018practical}
T.~Hlavacek, A.~Herzberg, H.~Shulman, and M.~Waidner, ``Practical experience:
  Methodologies for measuring route origin validation,'' in \emph{2018 48th
  Annual IEEE/IFIP International Conference on Dependable Systems and Networks
  (DSN)}.\hskip 1em plus 0.5em minus 0.4em\relax IEEE, 2018, pp. 634--641.

\bibitem{rodday2021revisiting}
N.~Rodday, I.~Cunha, R.~Bush, E.~Katz-Bassett, G.~D. Rodosek, T.~C. Schmidt,
  and M.~W{\"a}hlisch, ``Revisiting rpki route origin validation on the data
  plane,'' in \emph{Proc. of Network Traffic Measurement and Analysis
  Conference (TMA), IFIP}, 2021.

\bibitem{gilad2017we}
Y.~Gilad, A.~Cohen, A.~Herzberg, M.~Schapira, and H.~Shulman, ``{Are We There
  Yet? On RPKI's Deployment and Security},'' in \emph{NDSS}, 2017.

\bibitem{reuter2018towards}
A.~Reuter, R.~Bush, I.~Cunha, E.~Katz-Bassett, T.~C. Schmidt, and
  M.~W{\"a}hlisch, ``Towards a rigorous methodology for measuring adoption of
  rpki route validation and filtering,'' \emph{ACM SIGCOMM Computer
  Communication Review}, vol.~48, no.~1, pp. 19--27, 2018.

\bibitem{cloudflare}
Cloudflare, ``Is bgp safe yet?'' \url{https://isbgpsafeyet.com/}, accessed:
  04.09.2022.

\bibitem{apnic}
\BIBentryALTinterwordspacing
G.~Huston, ``Measuring roas and rov,'' 2021. [Online]. Available:
  \url{https://stats.labs.apnic.net/rpki}
\BIBentrySTDinterwordspacing

\bibitem{testart2020filter}
C.~Testart, P.~Richter, A.~King, A.~Dainotti, and D.~Clark, ``To filter or not
  to filter: Measuring the benefits of registering in the rpki today,'' in
  \emph{International Conference on Passive and Active Network
  Measurement}.\hskip 1em plus 0.5em minus 0.4em\relax Springer, 2020, pp.
  71--87.

\bibitem{DBLP:conf/uss/Birge-LeeSERM18}
\BIBentryALTinterwordspacing
H.~Birge{-}Lee, Y.~Sun, A.~Edmundson, J.~Rexford, and P.~Mittal, ``Bamboozling
  certificate authorities with {BGP},'' in \emph{27th {USENIX} Security
  Symposium, {USENIX} Security 2018, Baltimore, MD, USA, August 15-17, 2018},
  W.~Enck and A.~P. Felt, Eds.\hskip 1em plus 0.5em minus 0.4em\relax {USENIX}
  Association, 2018, pp. 833--849. [Online]. Available:
  \url{https://www.usenix.org/conference/usenixsecurity18/presentation/birge-lee}
\BIBentrySTDinterwordspacing

\bibitem{DBLP:journals/corr/abs-2004-09063}
\BIBentryALTinterwordspacing
Y.~Sun, M.~Apostolaki, H.~Birge{-}Lee, L.~Vanbever, J.~Rexford, M.~Chiang, and
  P.~Mittal, ``Securing internet applications from routing attacks,''
  \emph{CoRR}, vol. abs/2004.09063, 2020. [Online]. Available:
  \url{https://arxiv.org/abs/2004.09063}
\BIBentrySTDinterwordspacing

\bibitem{mitm:threat}
Renesys, ``{The New Threat: Targeted Internet Traffic Misdirection},''
  \\\url{http://www.renesys.com/2013/11/mitm-internet-hijacking/}, 2013.

\bibitem{indosat:hijack}
A.~Toonk, ``{Hijack Event Today by Indosat},''
  \\\url{http://www.bgpmon.net/hijack-event-today-by-indosat/}, 2014.

\bibitem{chung2019rpki}
T.~Chung, E.~Aben, T.~Bruijnzeels, B.~Chandrasekaran, D.~Choffnes, D.~Levin,
  B.~M. Maggs, A.~Mislove, R.~v. Rijswijk-Deij, J.~Rula \emph{et~al.}, ``Rpki
  is coming of age: A longitudinal study of rpki deployment and invalid route
  origins,'' in \emph{Proceedings of the Internet Measurement Conference},
  2019, pp. 406--419.

\bibitem{DBLP:conf/esorics/HlavacekSW22}
T.~Hlavacek, H.~Shulman, and M.~Waidner, ``Smart {RPKI} validation: Avoiding
  errors and preventing hijacks,'' in \emph{Computer Security - {ESORICS} 2022
  - 27th European Symposium on Research in Computer Security, Copenhagen,
  Denmark, September 26-30, 2022, Proceedings, Part {I}}, ser. Lecture Notes in
  Computer Science, V.~Atluri, R.~D. Pietro, C.~D. Jensen, and W.~Meng, Eds.,
  vol. 13554.\hskip 1em plus 0.5em minus 0.4em\relax Springer, 2022, pp.
  509--530.

\bibitem{DBLP:conf/ccs/HlavacekJMSW22}
T.~Hlavacek, P.~Jeitner, D.~Mirdita, H.~Shulman, and M.~Waidner, ``Behind the
  scenes of {RPKI},'' in \emph{Proceedings of the 2022 {ACM} {SIGSAC}
  Conference on Computer and Communications Security, {CCS} 2022, Los Angeles,
  CA, USA, November 7-11, 2022}, H.~Yin, A.~Stavrou, C.~Cremers, and E.~Shi,
  Eds.\hskip 1em plus 0.5em minus 0.4em\relax {ACM}, 2022, pp. 1413--1426.

\bibitem{usenix-stalloris-21}
\BIBentryALTinterwordspacing
``Stalloris: {RPKI} downgrade attack,'' in \emph{31st USENIX Security Symposium
  (USENIX Security 22)}.\hskip 1em plus 0.5em minus 0.4em\relax Boston, MA:
  USENIX Association, Aug. 2022. [Online]. Available:
  \url{https://www.usenix.org/conference/usenixsecurity22/presentation/hlavacek}
\BIBentrySTDinterwordspacing

\bibitem{DBLP:conf/ccs/MirditaSW22}
D.~Mirdita, H.~Shulman, and M.~Waidner, ``Poster: {RPKI} kill switch,'' in
  \emph{Proceedings of the 2022 {ACM} {SIGSAC} Conference on Computer and
  Communications Security, {CCS} 2022, Los Angeles, CA, USA, November 7-11,
  2022}, H.~Yin, A.~Stavrou, C.~Cremers, and E.~Shi, Eds.\hskip 1em plus 0.5em
  minus 0.4em\relax {ACM}, 2022, pp. 3423--3425.

\bibitem{hlavacek2020disco}
T.~Hlavacek, I.~Cunha, Y.~Gilad, A.~Herzberg, E.~Katz-Bassett, M.~Schapira, and
  H.~Shulman, ``Disco: Sidestepping rpki's deployment barriers,'' in
  \emph{Network and Distributed System Security Symposium (NDSS)}, 2020.

\bibitem{pearce2017augur}
P.~Pearce, R.~Ensafi, F.~Li, N.~Feamster, and V.~Paxson, ``Augur: Internet-wide
  detection of connectivity disruptions,'' in \emph{2017 IEEE Symposium on
  Security and Privacy (SP)}.\hskip 1em plus 0.5em minus 0.4em\relax IEEE,
  2017, pp. 427--443.

\bibitem{dai2021smap}
T.~Dai and H.~Shulman, ``Smap: Internet-wide scanning for spoofing,'' in
  \emph{Annual Computer Security Applications Conference}, 2021, pp.
  1039--1050.

\bibitem{DBLP:conf/dsn/ShulmanZ21}
H.~Shulman and S.~Zhao, ``Machine learning analysis of {IP} {ID}
  applications,'' in \emph{51st Annual {IEEE/IFIP} International Conference on
  Dependable Systems and Networks, {DSN} 2021, Taipei, Taiwan, June 21-24, 2021
  - Supplemental Volume}.\hskip 1em plus 0.5em minus 0.4em\relax {IEEE}, 2021,
  pp. 15--16.

\bibitem{zhang2018onis}
X.~Zhang, J.~Knockel, and J.~R. Crandall, ``Onis: Inferring tcp/ip-based trust
  relationships completely off-path,'' in \emph{IEEE INFOCOM 2018-IEEE
  Conference on Computer Communications}.\hskip 1em plus 0.5em minus
  0.4em\relax IEEE, 2018, pp. 2069--2077.

\bibitem{routeviews}
Routeviews, ``Routeview bgp collectors,''
  \url{http://www.routeviews.org/routeviews/index.php/collectors/}, accessed:
  25.07.2022.

\bibitem{RIS}
RIPE, ``Ris bgp collectors,''
  \url{https://www.ripe.net/analyse/internet-measurements/routing-information-service-ris/archive/ris-raw-data},
  accessed: 25.07.2022.

\bibitem{caidaASMapping}
CAIDA, ``Prefix to as mapping,''
  \url{https://www.caida.org/catalog/datasets/routeviews-prefix2as/}, accessed:
  25.07.2022.

\bibitem{caidaIXPMapping}
------, ``Ixp dataset,'' \url{https://www.caida.org/catalog/datasets/ixps/},
  accessed: 25.07.2022.

\bibitem{eqRPKI}
Equinix, ``Equinix rpki,''
  \url{https://docs.equinix.com/en-us/Content/Interconnection/IX/IX-rpki.htm},
  accessed: 30.07.2022.

\bibitem{aixRPKI}
AmsIX, ``Amsterdam ix routeservers,''
  \url{https://www.ams-ix.net/ams/documentation/ams-ix-route-servers},
  accessed: 30.07.2022.

\bibitem{mixRPKI}
MIX, ``Milan ix rpki,'' \url{https://www.mix-it.net/en/route-server/},
  accessed: 30.07.2022.

\bibitem{nixRPKI}
NIX.CZ, ``Nix.cz peering policy,''
  \url{https://www.nix.cz/file/PEERING_POLICY}, accessed: 30.07.2022.

\bibitem{decixRPKI}
DE-CIX, ``Routeserver guide rpki,''
  \url{https://www.de-cix.net/en/resources/service-information/route-server-guides/rpki},
  accessed: 30.07.2022.

\bibitem{cohen2001breakdown}
R.~Cohen, K.~Erez, D.~Ben-Avraham, and S.~Havlin, ``Breakdown of the internet
  under intentional attack,'' \emph{Physical review letters}, vol.~86, no.~16,
  p. 3682, 2001.

\bibitem{austein2017rfc}
R.~Austein, S.~Bellovin, R.~Housley, S.~Kent, W.~Kumari, D.~Montgomery,
  C.~Morrow, S.~Murphy, K.~Patel, J.~Scudder \emph{et~al.}, ``Rfc 8205-bgpsec
  protocol specification,'' 2017.

\end{thebibliography}
}

\end{document}